%% file: phi3.tex
\documentclass[prd,eqsecnum,12pt,tightenlines,nofootinbib,showpacs]{revtex4}
\usepackage{bm,graphicx,psfrag}

\newcommand{\JCCvec}[1]{\bm{#1}}

\newcommand{\MSbar}{\ensuremath{\overline{{\rm MS}}}}

\newtheorem{theorem}{%
   Theorem%
}
\newtheorem{definition}[theorem]{%
   Definition%
}
\newenvironment{proof}%
   {\begin{quotation}%
    \noindent
    {\bf Proof:}
   }
   {\end{quotation}}

\begin{document}

\preprint{
   \begin{tabular}{l}
     hep-ph/0110113 \\
     PSU/TH/211
   \end{tabular}
}
\title{Monte-Carlo Event Generators at NLO}

\author{John Collins}
\email{collins@phys.psu.edu}

\affiliation{
        Physics Department,
        Penn State University,
        104 Davey Laboratory,
        University Park PA 16802,
        U.S.A.
}

\date{8 October 2001}

\begin{abstract}%
  A method to construct Monte-Carlo event generators at arbitrarily
  non-leading order is explained for the case of a non-gauge theory.  A
  precise and correct treatment of parton kinematics is provided.
  Modifications of the conventional formalism are required: parton
  showering is not exactly the same as DGLAP evolution, and the external
  line prescription for the hard scattering differs from the LSZ
  prescription.  The prospects for extending the results to QCD are
  discussed.
\end{abstract}

\maketitle

\input{phi3-intro}
\input{phi3-main-features}
\input{phi3-definition}
\input{phi3-kinematics}
\input{phi3-examples}
\input{phi3-proof}
\input{phi3-algorithm}
\input{phi3-calculations}
\input{phi3-conclusions}

\begin{acknowledgments}
This work was supported in part by the U.S.\ Department of Energy
under grant number DE-FG02-90ER-40577.
I would like to thank DESY and the University of Hamburg for their
hospitality, and the Alexander von Humboldt foundation for an award 
during which this paper was completed.
I would like to thank Gunnar Ingelman, Torbj{\"o}rn Sj{\"o}strand and Dieter
Zeppenfeld for useful conversations.
\end{acknowledgments}


\end{document}

%% file: phi3-intro.tex

\section{Introduction}
\label{sec:intro}

There are two contrasting approaches to making perturbative predictions
from QCD.  The first consists of the analytic (or matrix element) methods
\cite{Analytic.Review}, where a factorization theorem is used to write
inclusive cross sections in terms of perturbatively calculable
hard-scattering coefficients and of parton densities and fragmentation
functions.  The second approach is that of the Monte-Carlo event generator
\cite{MC.Review}.  Now, the analytic methods have the advantage of
permitting calculations to arbitrary non-leading order, and thus of
allowing a systematic improvement in the accuracy of calculations.  In
contrast, the Monte-Carlo methods have the advantage of generating
complete events and of implementing QCD predictions for the detailed
structure of the final state, but at the expense of model dependence in
the hadronization.  But up to now, a systematic method of incorporating
higher order corrections, in powers of $\alpha_s$, has not been available
for the Monte-Carlo method.  Currently available event generators
incorporate hard scattering and evolution at the leading order
(leading-logarithm approximation, and somewhat beyond), and, although some
implementations of non-leading corrections have been made
\cite{MC.reweighting,MC.NLO.subtractive}, there has not so far been
devised a general method.

It is of course highly desirable to extend the algorithms in the event
generators to non-leading order.  Once this aim has been achieved, one
can use the highest precision QCD calculations while retaining the
advantages of the event generators.  The many currently available NLO
and NNLO analytic calculations of Feynman graphs will of course
continue to be used; they will merely need to be reformulated to be
used in event generators.

In Ref.\ \cite{MC.BGF}, I proposed a subtractive approach and applied
it to the simple, but phenomenologically important, process of
photon-gluon-fusion in deep inelastic scattering.  This approach was
intended to be generalized to provide a complete solution to the
problem of non-leading corrections to event generators.
Unfortunately, a full treatment of QCD is quite hard, because of the
need to treat soft gluon effects.  So the present paper restricts its
attention to providing a complete treatment within a model theory
where only collinear configurations are important.

For ease of exposition, only the rather unphysical case of $\phi^3$
theory in 6 space-time dimensions will be discussed, but the methods
immediately generalize to any non-gauge theory\footnote{ 
    In a theory with fermionic fields, for example, spin effects
    can be included by the method I proposed in \cite{MC.spin}.  
}.  
In such a theory, we will see in complete generality how to construct
Monte-Carlo event generators for $e^+e^-$ annihilation with the inclusion
of arbitrarily non-leading order corrections, in the hard scattering and
in the final-state showering.  The irreducible errors are then due to
power-law corrections to the hard scattering and evolution kernels.  The
extension to initial-state showering should be relatively elementary.

The general principle of the method is to generate extra kinds of hard
interaction and splitting kernels corresponding to non-leading-order
corrections, and to apply subtractions to allow for the effect of
lower-order terms coupled to showering.  The subtractions are applied
point-by-point in momentum space with proper allowance for correct parton
kinematics, so that the NLO corrections form completely well-behaved
functions.  This is in contrast to the subtractive methods used in
parton-level Monte-Carlo calculations for infra-red-safe observables, as
in Ref.\ \cite{MC.analytic}.  In those methods the NLO corrections are
singular distributions which can only be used to calculate suitably
infra-red-safe observables.  The new method is applicable to general cross
sections.

The viewpoint in this paper can be characterized as ``static'': It
asks what the value is of a certain sum of integrals for the cross
section.  This contrasts with the more dynamic viewpoint normally used
in discussions of event generators, where the primary concept is the
evolution of partonic states.  An actual implementation of the methods
of this paper will recover this dynamic point-of-view, for that is
inherent in the structure of the Monte-Carlo algorithms derived from
the static Feynman graph structure.

Although the model theory treated in this paper is relatively simple,
even this case already poses some quite non-trivial difficulties if
one is to solve completely and reliably the problem of obtaining
corrections for use in an event generator.  Arbitrarily high-order
Feynman graphs are being approximated in such a way that an estimate
of the cross section for producing $N$ particles needs computational
resources proportional to $N$, even though a brute force calculation
requires resources more like $N!$.  

Since the mathematical structure worked out in this paper is quite
general, I hope that the same structure and proof will apply in many
situations, with minor modifications.  The essential problem to be
solved in this and any other problem in asymptotic behavior in quantum
field theories (including QCD) is to find an economical way of
treating the large multiplicity of different regions that give the
leading contributions from a single graph.  Any economical method must
treat exactly one region at a time, and involve only simple and very
general relations to other regions.

In addition, the methods used in the proof should be of use in algorithms
for QCD calculations (both numeric and symbolic).  Because of the many
singularities in unintegrated graphs, it is important to have robust
control over the sizes of the contributions from the different regions,
and to have a very concrete definition of subtracted integrands, which are
to be well-behaved numerically.  The methods of proof in this paper, with
their simple but strongly recursive structure, are appropriate for
generalization for this purpose.

In Sec.\ \ref{sec:new.features} are summarized some of the general
ideas used in this paper.  Then, in Sec.\ \ref{sec:basics}, we define
the cross section that is to be treated and construct the basic
elements that are used in the algorithm.  The basic ingredient used to
construct the factors in the factorization theorem is an approximation
where the external lines of the hard-scattering are set on-shell.
Precise definitions of this approximation are given in Sec.\ 
\ref{sec:kinematics}.  The definitions are arranged to be suitable for
a Monte-Carlo event generator, where it is essential that parton
4-momentum is exactly conserved; approximations that change the total
4-momentum of the scattering have to be used carefully.  Before giving
the general proofs, we show, in Sec.\ \ref{sec:examples}, how the
methods are applied to specific Feynman graphs.  In Sec.\ 
\ref{sec:factorization.proof}, a general proof of factorization is
given in the form that it is needed to obtain an algorithm for a
Monte-Carlo event generator.  This section also includes a proof of
the collinear safety of the hard scattering and the integrated jet
factors, as well as a discussion of the renormalization group
properties of the factors in the factorization theorem.  These are all
essential to an actual implementation of the algorithm.  In Sec.\ 
\ref{sec:monte.carlo.algorithm}, we then obtain the Monte-Carlo
algorithm; this is the culmination of the theoretical treatment.
Sec.\ \ref{sec:calculation.examples} shows examples of complete
calculations of corrections to the hard-scattering, at NLO.  Finally,
some conclusions are presented in Sec.\ \ref{sec:conclusions} and some
of the difficulties that remain in extending the method to QCD are
explained.


%% file: phi3-main-features.tex

\section{Main features of the new method}
\label{sec:new.features}

\subsection{Previous methods}

\begin{figure}
\includegraphics[scale=0.8]{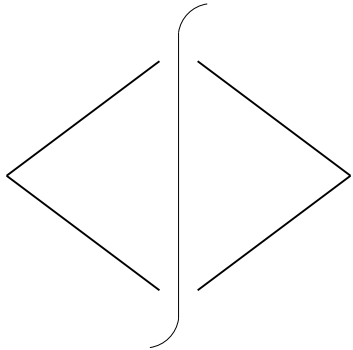}
\caption{$e^+e^-$ annihilation to 2 partons in lowest order.}
\label{fig:2jet}
\end{figure}

\begin{figure}
\psfrag{p_1}{$p_1$}
\psfrag{p_2}{$p_2$}
\psfrag{p_3}{$p_3$}
\includegraphics[scale=0.8]{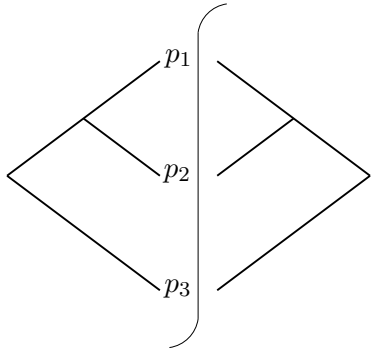}
\caption{A graph for $e^+e^-$ annihilation to 3 partons.}
\label{fig:3jet}
\end{figure}

The methods used to treat higher-order corrections in the analytic
methods, even within a Monte-Carlo approach \cite{MC.analytic}, do not
directly translate to methods for event generators.  One of the reasons is
the way in which approximations are made on the kinematics of events.
Consider, for example, the production of two or three partons (in
$e^+e^-$-annihilation). The lowest order graph for 2-parton production,
shown in Fig.\ \ref{fig:2jet}, is assigned to an exactly back-to-back
configuration.  In the lowest order graph for 3-parton production, Fig.\ 
\ref{fig:3jet}, the 3 partons are not back-to-back.  However there is a
singularity in the 3-parton calculation when pairs of the final-state
particles approach each other and make a 2-jet configuration.  This part
of the cross-section is really a 2-jet process with an order $\alpha_s$
contribution to the fragmentation of one of the jets.

In the standard calculation of the hard-scattering coefficient for
3-parton configurations, a subtraction is made of a fragmentation factor
times the 2-parton cross section, which is localized in exactly the
back-to-back situation. Let $\JCCvec{q}_T$ be a transverse momentum that
parameterizes the deviation from a back-to-back configuration.  Then the
resulting contribution to the hard-scattering coefficient for the 3-jet
cross section is of the form:
\begin{equation}
\label{eq:qT.distribution}
    \left( \frac {a \ln q_T^2 + b}{q_T^2} \right)_+
    + c \, \delta^{(2)}(\JCCvec{q}_T) ,
\end{equation}
plus some irrelevant non-singular terms.  Without the $+$ prescription,
the cross section has a divergent integral over $\JCCvec{q}_T$.  Although
the subtraction at $q_T=0$ produces a valid, integrable distribution (or
generalized function), it does not produce an ordinary function.  To
implement a 3-jet cross section in an event generator one should have an
ordinary function, not a distribution.

A related difficulty is that after showering and hadronization of the
final-state partons, the two graphs, Figs.\ \ref{fig:2jet} and
\ref{fig:3jet}, produce hadrons in overlapping regions: one cannot
distinguish between an event that is derived from exactly back-to-back
partons, as in Fig.\ \ref{fig:2jet}, and an event derived from an almost
back-to-back configuration from Fig.\ \ref{fig:3jet}.  However, the
distribution in Eq.\ (\ref{eq:qT.distribution}) does make such a
distinction.  Moreover from the large $q_T$ tail of the hadronization of
the 2-jet cross section one obtains final states with 3 (or more jets).

The simplest solution to these problems is to compute an infra-red
safe observable, like thrust.  This is commonly done with jet
production in $e^+e^-$ annihilation \cite{MC.analytic}.  But this is
not suitable in an event generator, where exclusive final states are
generated and arbitrary observables can be estimated. 

Another method is commonly used in hadron-hadron collisions, where the
cross sections for all jet observables involve parton densities, so that
purely perturbative calculations are never IR safe.  This method is to
convolute Eq.\ (\ref{eq:qT.distribution}) with some kind of smearing
function, that intuitively is supposed to represent intrinsic transverse
momentum of partons relative to corresponding hadrons.  Without further
justification, this is just an ad hoc prescription that is used in a
attempt to model a physical phenomenon that is clearly present.
Furthermore, some of the $q_T$ smearing is contained in higher-order
corrections to the hard scattering.  To implement a valid calculation, it
is necessary to devise a subtraction method that is derived from the
underlying field theory and that prevents double counting; in a sense this
gives a correct method to implement $q_T$ smearing.  

One very different approach that has been used \cite{MC.reweighting} to
implement NLO corrections in event generators is to reweight events after
they have been generated, so that the hard scattering cross section is
renormalized to the value known from analytic calculations.  In addition,
the unsubtracted NLO correction is used to populate regions of phase space
not filled by showering the partons associated with the LO cross section.
This method does not readily generalize.

\subsection{Overall structure of new method}

\begin{figure}
\begin{displaymath}
\psfrag{l_1}{$l_1$}
\psfrag{l_N}{$l_N$}
\psfrag{J_1}{$J_1$}
\psfrag{J_N}{$J_N$}
\psfrag{H}{$\!H$}
  \sigma = \sum_{N=2}^\infty ~ ~ ~ \raisebox{-0.5\height}{\includegraphics{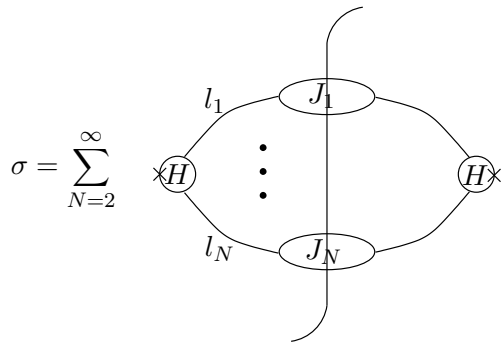}}
\end{displaymath}
\caption{General leading-power contribution to cross section}
\label{fig:Njet}
\end{figure}

The basic idea of the new method is that the subtractions should be
computed with the correct parton kinematics, rather than a
zero-transverse-momentum approximation such as is implicit in the
$+$-distribution in Eq.\ (\ref{eq:qT.distribution}).  This means that we
must always remember that graphs like Figs.\ \ref{fig:2jet} and
\ref{fig:3jet} do not fully represent the physics.  The parton lines going
into the final state should really be replaced by bubbles representing
hadronization, as in Fig.\ \ref{fig:Njet}, which represents the most
general leading region.

Apart from such implementation issues, the basis of the method is the
usual factorization theorem: As illustrated in Fig.\ \ref{fig:Njet},
the leading power contributions involve a hard scattering convoluted
with jets, each jet being induced by a single parton of relatively low
virtuality.  The algorithm for the event generator makes recursive use
of the factorization theorem by applying it to each of the jets
whenever the invariant mass of the parton initiating the jet is much
larger than a typical hadronic mass.  In this second application of
factorization to the jets, the hard-scattering coefficients are
equivalent to the DGLAP splitting kernels.

Current event generators are based on a leading-order approximation
which involves the use of the lowest order expansion of the
hard-scattering and splitting kernels together with Sudakov form
factors that encode the renormalization-group transformation between
the potentially widely different scale in the different applications
of factorization.

What has been gained by the use of factorization is the ability to
compute an arbitrarily large number of high-order Feynman graphs in a
time that is merely proportional to the number of particles in the
final state.  The price of this is that approximations have been made,
both as regards the kinematics of individual graphs and as regards the
subset of graphs that is calculated.  It is the proof of the
factorization theorem that shows that the approximations are valid and
useful. 

At each stage of the algorithm, there are errors due to higher-order
corrections to the relevant hard scattering.  The higher-order
contributions to the cross section are computed by subtracting from the
bare graphs contributing to the $H$ subgraphs in Fig.\ \ref{fig:Njet} the
contributions at the same order that are allowed for in the lower-order
calculation combined with showering.  To devise an algorithm one must: (a)
define precisely what is being computed, (b) define precisely the
kinematic approximations used in the hard scattering, and (c) implement
the subtractions point-by-point in momentum space.  The subtraction terms
are computed with the same kinematics as the bare term, to satisfy the
last requirement, while the kinematic approximations are only used in the
computation of the numerical values of the matrix elements, but not in
defining the parton momenta.  A similar treatment of initial-state
showering will require us to employ parton densities differential in
transverse momentum, or some equivalent concept.

Once the method is formulated properly for the higher order
corrections to the hard scattering, it is readily extended to the
evolution kernels, as we will see.

One potential problem is that, as is well known, subtraction methods
tend to generate terms in the cross section that are negative.  The
reason is that if, as is possible, a next-to-leading (NLO) correction
decreases the cross section compared with the leading-order (LO) cross
section, then the NLO term is negative.  The virtual corrections to
the hard scattering can simply be added to the LO term; in the method
proposed in this paper, these virtual corrections will always be
collinear and infra-red safe, even in QCD --- see the work of Collins
and Hautmann \cite{Collins:2000dz,Collins:2000gd} --- since the
collinear and soft pieces are supposed to subtracted.  Thus the
virtual corrections cause no problems of positivity.

As for the real NLO corrections, they are treated as generating a
separate class of events.  Where their contribution to the
differential cross section is negative, these events must be generated
with negative weight.  This is often considered undesirable, but as
explained in \cite{MC.BGF}, this is actually acceptable for most
purposes, provided that only a reasonably small number of events with
moderate negative weights is generated.  Then the negative-weight
events are outweighed by larger positive contributions to a binned
cross section.  What is not acceptable is to have the cross section be
obtained as a cancellation of relatively much larger positive and
negative contributions.  An extreme example of this is given by Eq.\ 
(\ref{eq:qT.distribution}).  There the distribution $(1/q_T^2)_+$
consists of an unsubtracted $1/q_T^2$ term, which integrates to
positive infinity, and a delta function with an infinite negative
coefficient.  It is also essential that all regions of final-state
momentum space that are populated by negative-weight NLO events are
also populated by positive-weight LO events; thus the cancellations
that give a positive physical cross section occur point-by-point
rather than between neighboring bins.  Both these conditions are
fulfilled by my algorithm.

In addition, the Monte-Carlo formalism contains some cut-off
functions.  When the higher-order corrections are included, there is a
renormalization-group invariance under changes of the cut-off
functions, and these can be adjusted \cite{MC.BGF} to at least reduce
the number of negative-weight events. 


%% file: phi3-definition.tex

\section{Definition of problem}
\label{sec:basics}

\subsection{Model for $e^+e^-$ annihilation}
\label{sec:model}

Let us use $\phi^3$ theory in 6 space-time dimensions, which has the
Lagrangian
\begin{equation}
   {\cal L} = \frac {1}{2} \partial\phi^2
              - \frac {m^2}{2} \phi^2
              - \frac {g}{6} \phi^3.
\end{equation}
In addition, to mimic an interaction with a photon, let us suppose
that there is a weakly interacting scalar field $A$, which will be
called a photon, and that the interaction with the $\phi$ field is
proportional to $Aj$ where the ``current'' $j$ is the renormalized
operator\footnote{
   The square bracket notation used in this context denotes a
   renormalized operator.  
}
$[\frac {1}{2}\phi^2]$.  To emphasize the analogy that is intended
with QCD, let us call $\phi$ the ``parton'' field.

The process we work with is an analog of $e^+e^-$ annihilation to
hadrons.  It will be necessary to consider various cross sections
(including, for example, the total cross section, jet cross sections
and completely exclusive cross sections), so to unify the treatment I
define a weighted cross section by
\begin{equation}
\label{eq:sigma.W}
   \sigma [W]
    = K
      \sum_f W(f)
      \int d^6x \, e^{iq\cdot x}
      \langle 0 | j(x) |f\rangle \langle f | j(0) | 0 \rangle .
\end{equation}
Here $q^\mu$ is the total incoming momentum, $j(x)$ is the current operator
defined above, and $K$ is a standard factor that depends only on the
leptonic variables.  The sum over $f$ denotes an integral and sum over all
final states \footnote{
   It is probably most useful to treat the sum and integral over final
   states as being over all states without regard to the constraints
   imposed by indistinguishability.  Then the overcounting of physically
   identical states is to be compensated by inserting a suitable factor,
   which for our model theory is $1/N_f!$, where $N_f$ is the number of 
   final-state particles.  
}, and $W(f)$ is a weighting function that defines the particular
cross section under consideration.  All cross sections can be obtained 
if $\sigma[W]$ is known for all $W$. Let us work in the overall
center-of-mass frame, $q^\mu = (Q, \JCCvec{0})$. 

Examples of the choice of the weighting function are:
\begin{itemize}

\item
   The total cross section, given by $W=1$ for all final states. 

\item
   The 2-jet cross section, for which $W(f)=1$ for all final states
   containing exactly two jets (given some definition of a jet), and
   $W(f)=0$ otherwise.

\item
   A fully differential two particle cross section, for which $W(f)$
   is the appropriate $\delta$ function for two particle final states, and 
   for which $W(f)=0$ for final states containing three or more
   particles. 

\end{itemize}

\subsection{Leading-power contributions}
\label{sec:leading.power}

Our aim is to decompose the cross section into its completely
exclusive components, and to obtain an algorithm for calculating them
to the leading power in $Q$.  Now, the usual power-counting theorems
of Libby and Sterman \cite{pc}, which are formulated in the overall
center-of-mass frame, show that the leading-power contributions to the
cross section in our model theory all come from regions of the form
symbolized in Fig.\ \ref{fig:Njet}.  Some lines, forming the hard
subgraph $H$ and its complex conjugate, are off shell by order $Q^2$.
The other lines form groups, or jets, of lines of approximately
parallel momentum.  These groups are labeled $J_{1}$ \dots $J_{N}$,
and to satisfy momentum conservation there must be at least two jets,
$N \geq 2$.

Unlike the case of QCD (or any other gauge theory), there are no soft
subgraphs in our model $\phi^3$ theory, at the leading power.  This fact
alone considerably simplifies the analysis in our model compared with QCD.

\begin{figure}
\psfrag{p_1}{$p_1$}
\psfrag{p_2}{$p_2$}
\psfrag{p_3}{$p_3$}
\psfrag{p_4}{$p_4$}
\includegraphics{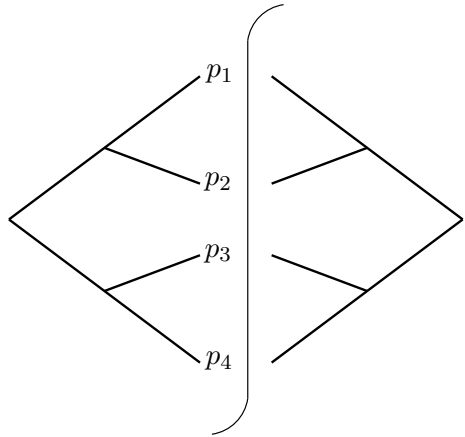}
\caption{Graph with several leading regions}
\label{fig:4jet}
\end{figure}

Individual Feynman graphs for the cross section can contribute in
several different regions.  For example, consider the cut graph, $\Gamma$,
of Fig.\ \ref{fig:4jet}, which has four particles in the final state,
of momenta $p_1$, $p_2$, $p_3$ and $p_4$.  The graph has the following
regions for leading twist contributions:
\begin{enumerate}

\item
    $p_1$ and $p_2$ are almost parallel, and $p_3$ and $p_4$ are also
    almost parallel.

\item
    $p_3$ and $p_4$ are almost parallel, but $p_1$ and $p_2$ are at
    wide angle.

\item
    $p_1$ and $p_2$ are almost parallel, but $p_3$ and $p_4$ are at
    wide angle.

\item
    All the final-state momenta are at wide angles with respect to
    each other.  

\end{enumerate}
There are of course contributions from intermediate regions, and these 
give the leading-logarithmic part of the cross section.  The
techniques of our proof will avoid any need to consider these
intermediate regions explicitly.

\subsection{Summary of proof of factorization and application to event
  generator} 
\label{sec:proof.summary}

A summary of the proof of the factorization theorem in perturbation theory
is as follows:
\begin{enumerate}

\item Write the cross section as a sum over graphs $\Gamma$.
    
\item Write each graph as a sum of terms $C_R(\Gamma)$, one for each
    of its leading regions $R$:
    \begin{equation}
    \label{eq:sum.over.regions}
       \Gamma = \sum_{{\rm regions~}R} C_R(\Gamma) 
       + \mbox{non-leading power of $Q$} .
    \end{equation}

\item Sum over all possibilities:
    \begin{eqnarray}
    \label{eq:basis.of.proof}
       \sigma[W] &=& K \sum_{{\rm final~states~}f} 
                     ~ \sum_{{\rm graphs~}\Gamma} \Gamma \, W(f) 
    \nonumber\\
       &=& K \sum_f \sum_{\Gamma,R} C_R(\Gamma) W(f)
         +  \mbox{non-leading power of $Q$} .
    \end{eqnarray}
    Here $K$ is the same overall factor that we used in Eq.\
    (\ref{eq:sigma.W}). 
    
\item Now $C_R(\Gamma)$ (to be defined later) is the product of a hard
    part and a set of jet factors, with these factors corresponding to 
    particular contributions to the subgraphs $H$ and $J_i$ in Fig.\
    \ref{fig:Njet}.  The sum over graphs and regions can be reformulated
    as independent sums over the hard part and the jet subgraphs, and this
    gives the desired factorization theorem: 
    \begin{equation}
    \label{eq:basic.factorization}
      \sigma[W] = K \sum_f \sum_N \hat H_N \times \Delta 
                     \times \prod_{j=1}^N J_j \, W(f)
             +  \mbox{non-leading power} .
    \end{equation}
    Here, the hard part is denoted by $\hat H_N$, with a hat over the $H$.
    The hat indicates that the contributions to the hard part are not
    obtained simply as the contributions of appropriate Feynman graphs in
    a certain region of momentum space.  Instead, as we will see, they are
    obtained from the relevant Feynman graphs with the aid of a
    subtraction procedure that prevents double counting.  The factor $\Delta$
    is a Jacobian, whose calculation we will see later; it relates the
    integration over the massless final-state momenta of $\hat{H}_N$ to
    the integration over the correct and off-shell parton momenta.  
    
\item It is convenient to adjust the normalizations to give a formula
    that has a convenient interpretation as a hard-scattering factor times
    jet factors that are normalized to be probabilities---see Eq.\ 
    (\ref{eq:factorization3}) below.

\end{enumerate}
In a Monte-Carlo event generator, all integrals over final states are
done by a Monte-Carlo technique.  After generating the distribution of
partons associated with the hard scattering, the distribution over
invariant mass for each jet-initiating parton $l_j$ is computed.
Because of the collinear safety of total cross sections, this is a
perturbative calculation when the virtuality of $l_j$ is large.  Then
a further application of the factorization theorem is made to each
jet, whenever the invariant mass of the initiating parton of the jet
is large enough for the use of perturbative methods.  This procedure
is applied recursively until all the generated partons have low
invariant mass.  After that a hadronization model is applied, so that
a hadronic final state is generated.  Finally, the weight function
$W(f)$ may be applied, to define a particular cross section.  Since
the weight function is only applied at the last stage, all the
previous steps can be implemented independently of the cross section
being computed.  Hence the events can in fact be generated without
regard to the weight function; the weight function simply provides a
convenient tool to unify all cross section calculations into a single
formula.

\subsection{Comments}

The proof is given only in perturbation theory, but with a treatment of
all orders at the leading power.  Nevertheless the structure that is
derived is evidently more general.  Thus one can conjecture that a better
and less perturbative treatment is possible.  In particular, it is very
natural that the structure of Fig.\ \ref{fig:Njet} should hold beyond
perturbation theory.

Factorization is needed so that the structure of the algorithm for
generating events is simple.
One important feature is that the hard scattering contribution from
each region is computed in the approximation that its external lines
have zero virtuality.  Thus the computation of the hard-scattering can
be done independently of the subsequent computation of the properties
of the jets.  This approximation is good up to the claimed power-law
correction when the virtualities of the jet-initiating partons are
low.  When the virtualities of these partons get large, the
approximation worsens.  But then, as we will see, the higher-order
corrections to the hard scattering correctly handle this situation, and
compensate the errors. 

For the decomposition of the graphs into a sum over regions,
phase-space slicing is the most obvious method.  However it does not
produce the desired result, since the errors are fairly hard to control,
and the errors tend to be quite non-optimal\footnote{
  The word ``error'' in this context means the difference between the
  exact value of a graph and the approximation used in the sum over
  regions. 
}.  This will lead us to use a more formal subtraction method.  

Essential requirements on its implementation appear to be:
\begin{itemize}
\item Each particular subgraph for $H$ and $J_j$ gives the same result
  in all contexts, independently of the graph $\Gamma$ from which the
  subgraph arises. This allows the manipulations between Eqs.\ 
  (\ref{eq:basis.of.proof}) and (\ref{eq:basic.factorization}) to work.
\item The space of final states factorizes into parts associated with
  the partonic final states of the hard scattering and parts
  associated with the final states of each jet.  Then factorization
  for the cross section follows from factorization for the amplitudes
  in each region.  Again this is needed so that the manipulations between Eqs.\ 
  (\ref{eq:basis.of.proof}) and (\ref{eq:basic.factorization}) work.
\item The quantity $C_R(\Gamma)$ that defines the contribution of a graph
  $\Gamma$ in a region $R$ must be an ordinary function, not a singular
  distribution.  Then the integral over final states can be
  implemented by a Monte-Carlo method.
\end{itemize}

\subsection{Errors}

It is a little complicated to explain what are the errors on the
approximations we make.  This is an important issue to get correct,
because if an event generator is used to make predictions that are
compared with experiment, we need to understand the significance of
deviations between the data and the predictions.

First observe that the amplitudes are treated with the correct final
states but with an approximation on the values of the amplitudes.  The
power counting theorems of Libby and Sterman show that the total
cross-section is computed up to an error that is suppressed by a power of
$Q$.  Moreover this cross section is correctly decomposed into its
exclusive components.  So the obvious expectation is that the differential
Monte-Carlo cross section differs from the exact cross section by a power
suppressed error:
\begin{equation}
\label{eq:bad.error}
  \sigma_{\rm exact}[W] - \sigma_{\rm calculated}[W]
  = \sigma_{\rm exact}[W] \times O(m/Q)^p 
~\mbox{(incorrect estimate)}.
\end{equation}
Here $m$ is a typical hadronic mass scale, and $p$ is some positive
number.  For most final states, this estimate is indeed correct, since the
approximations used are good to power-law accuracy in each leading region.

However, there are other, non-leading regions, and for the theories
under consideration this means regions that are not of the form of
Fig.\ \ref{fig:Njet}.  The approximations we made are not valid in
these regions, so a substantial deviation between theory and
experiment can occur in a non-leading region.  On the other hand, a
non-leading region gives a power-suppressed contribution to the cross
section; that is exactly what it means to say that the region is
non-leading.  So typically we will have small contributions to the
cross section from such a region.

Therefore we need to modify the error estimate.  The following is one
possibility: 
\begin{equation}
\label{eq:weak.error}
  \sigma_{\rm exact}[W] - \sigma_{\rm calculated}[W]
  = \sigma_{\rm total} \times O(m/Q)^p 
~\mbox{(weak estimate)} .
\end{equation}
This says that the error is power suppressed with respect to the total
cross section.  Unlike the previous estimate, it is excessively weak,
since the non-leading regions will populate only certain regions of the
space of final states.  The simplest correct estimate is
\begin{equation}
\label{eq:error}
  \sigma_{\rm exact}[W] - \sigma_{\rm calculated}[W]
  = \sigma_{\rm calculated}[W] \times O(m/Q)^p 
    + \mbox{terms for non-leading regions} .
\end{equation}
This says that the errors come from two sources: power-suppressed errors
in the approximations, and contributions from non-leading regions.  It
leaves open the issue of exactly which hadronic final states are populated
by the non-leading partonic regions.

\begin{figure}
\psfrag{p_1}{$p_1$}
\psfrag{p_2}{$p_2$}
\psfrag{p_3}{$p_3$}
\includegraphics[scale=0.8]{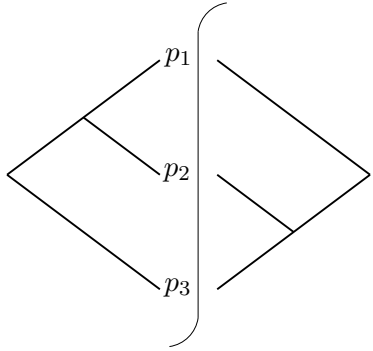}
\caption{A graph for $e^+e^-$ annihilation to 3 partons with no singular
   leading regions.}
\label{fig:non.lead}
\end{figure}

An example of the issues can be seen with the aid of the graph of Fig.\
\ref{fig:non.lead}.  The only leading region for this graph is the wide
angle region, where the partonic final state makes 3 jets.  It also has
non-leading regions.  Among these are 2-jet regions, where 2 partons are
close in angle.  Only one propagator has a small denominator, so the
graph's contribution is much smaller than the contribution of graphs like
those of Fig.\ \ref{fig:3jet}, where two denominators are small.  Since
there are other contributions to the cross section that are bigger, we see
that the graph makes a power-suppressed contribution to the 2-jet region.
In addition, there is a region where parton $p_2$ is soft, i.e., has all
its momentum components much less than $Q$.  This region would be leading
in QCD, if the soft parton were a gluon, but the region is non-leading in
a non-gauge theory.  However, the kinematic region it gives in the final
state is not necessarily populated by the leading regions, so this case
gives a non-trivial example of the last term on the right-hand side of
Eq.\ (\ref{eq:error}).


%% file: phi3-kinematics.tex

\section{Kinematics of hard scattering}
\label{sec:kinematics}

The core of the proof of factorization is a treatment of the sum over
leading regions for an individual graph in Eq.\ 
(\ref{eq:sum.over.regions}).  Each term $C_R(\Gamma)$ is defined by an
appropriate kinematic approximation and a subtraction procedure.  Given
this definition, one must prove the error estimate in Eq.\ 
(\ref{eq:sum.over.regions}).  Various properties of the kinematic
approximations and the subtractions are needed to obtain the Monte-Carlo
algorithm and the methods for the calculation of non-leading corrections.
These properties ensure that the combinatorial part of the proof, between
Eqs.\ (\ref{eq:basis.of.proof}) and (\ref{eq:basic.factorization}),
actually works.

In this section we define the approximation associated with a
particular region, and derive its important properties.

\subsection{Approximation}
\label{sec:approximation}

Consider a generic leading-power contribution from a particular graph
for the cross section, as symbolized in Fig.\ \ref{fig:Njet}.  From it
we will now construct an approximation that is suitable for obtaining
a factorization theorem and a Monte-Carlo implementation.  The main
idea of the approximation is to replace the external momenta of the
hard subgraph $H$ by on-shell massless momenta, and also to set the
masses of the internal propagators of $H$ to zero.  Since the
approximation will be integrated to a region outside of its domain of
validity, it will be necessary to give precise definitions of the
transformation from the exact momenta to the massless approximated
momenta and to compute the Jacobian of the transformation.

Let $\Gamma[W]$ denote the particular graph that is being computed, and let
it be decomposed into subgraphs corresponding to a particular region
$R$ of the form of Fig.\ \ref{fig:Njet}, with $H_R$ being the hard
subgraph and with $J_{R1}$ \dots $J_{N_R}$ being the jet
subgraphs:\footnote{
   Note that the symbols $H_R$ and $J_{Rj}$ are not to be identified with
   the corresponding symbols in Eq.\ (\ref{eq:basic.factorization}),
   although they have related meanings.  In that equation, $\hat{H}_N$ and
   $J_j$ mean the fully subtracted contributions to the the hard part
   and the jet factors, with sums over all contributing graphs.  In
   contrast, $H_R$ and $J_{Rj}$ in the current section mean the values of
   subgraphs of a particular graph $\Gamma$, with the subgraphs
   being the hard and jet subgraphs associated with a particular
   region $R$. 
}
\begin{equation}
\label{eq:Njet.region}
    \Gamma[W] = \int dL(p;q,m) \, H_R(l) \prod_j J_{Rj}(p_j) W(f) .
\end{equation}
Here the momenta are defined as follows:
\begin{itemize}

\item
    The final-state particles of jet subgraph $J_{Rj}$ are
    $p_{j,1}^\mu$, \dots, $p_{j,n_j}^\mu$, where $n_j$ is
    the number of final state lines of $J_{Rj}$.

\item
    Then $p_j$, with one subscript, denotes the collection of these
    $n_j$ momenta, i.e., $p_j^\mu = \left(p_{j,1}^\mu,
    \dots, p_{j,n_j}^\mu\right)$.
    The sum of these momenta is $l_j = \sum_{i=1}^{n_j} p_{j,i}$,
    which is the momentum of the line joining $J_{Rj}$ to the hard
    subgraph $H_R$.

\item
   The symbol $p$, with no subscripts denotes the collection of all
   momenta of the final state particles.

\item
    The symbol $l$, with no subscript, denotes the collection of the
    $l_j$'s, i.e., 
    $l^\mu = \left( l_1^\mu, \dots, l_{N_R}^\mu \right)$.  The sum of these
    momenta is the photon momentum: 
    $\sum_{i,j} p_{j,i}^\mu = \sum_j l_j^\mu = q^\mu$. 

\end{itemize}
The symbol $dL(p;q,m)$ denotes the measure for integration over the
Lorentz-invariant phase space:
\begin{equation}
    dL(p;q,m) = \prod_{i,j} \frac{ d^5\JCCvec{p}_{j,i} }{ (2\pi)^5 2E_{j,i} }
           \, (2\pi)^6 \, \delta^{(6)}\!( \sum_{i,j}p_{j,i} - q ) ,
\end{equation}
with the argument $m$ of $dL$ denoting the mass of each of the final-state
particles.

In Eq.\ (\ref{eq:Njet.region}), $H_R(l)$ denotes the hard factor, i.e., the
expression for the product of the two subgraphs labeled $H$ in Fig.\
\ref{fig:Njet} on each side of the final-state cut.  The hard subgraph is
to be amputated, i.e., it is without any factors for its external lines,
and it is to be integrated over all its internal loop momenta.

Similarly each $J_{Rj}(p_j)$ denotes the corresponding jet factor in Fig.\ 
\ref{fig:Njet}.  All its external line factors are to be included.  It is
also to be integrated over all its internal loop momenta.

It is important to the derivation of a factorization theorem that the
integration over final-state phase space in Eq.\ 
(\ref{eq:Njet.region}) can itself be factorized:
\begin{eqnarray}
\label{eq:phase.space}
   dL(p;q,m) &=&
    \left(\prod_j \int \frac{ d^6l_j }{ (2\pi)^6 }
    \right)
    \, (2\pi)^6 \, \delta^{(6)}\!( \sum_jl_j - q )
    \prod_j \left[
        \prod_{i} \frac{ d^5\JCCvec{p}_{j,i} }{ (2\pi)^5 2E_{j,i} }
        \, (2\pi)^6 \, \delta^{(6)}\!( \sum_{i}p_{j,i} - l_j )
    \right]
\nonumber\\
&=&
    \left(\prod_j \int \frac {d^6l_j}{(2\pi)^6}
    \right)
    \, (2\pi)^6 \, \delta^{(6)}\!(\sum_jl_j - q)
    \left( \prod_j dL(p_j;l_j,m) \right)
\nonumber\\
&=&
    \left(\prod_j  \int \frac {dM_j^2}{2\pi } \right)
    \left( \prod_j  
       \int \frac {d^5\JCCvec{l}_j}
            {(2\pi)^5 2\sqrt{\JCCvec{l}_j^2 + M_j^2}}
    \right)
    \, (2\pi)^6 \, \delta^{(6)}\!(\sum_jl_j - q)
    \left( \prod_j dL(p_j;l_j,m) \right)
\nonumber\\
&=&
    \left(\prod_j  \int \frac {dM_j^2}{2\pi } \right)
    ~\int dL(l;q,M)~
    \left( \prod_j dL(p_j;l_j,m) \right).
\end{eqnarray}
In the first and second lines the phase space is decomposed into
separate phase-space integrals $dL(p_j;l_j)$ for the final-state
momenta of each jet. Then in the third and fourth lines, the
integration over the $l_j$'s is decomposed according to the invariant
mass of these momenta, so that the integration is presented as an
integral over the momentum of a set of $N_R$ particles with masses
$M_1$, \dots, $M_{N_R}$.  Notice that in the last line it was
necessary to make explicit the mass arguments for the phase space of
the jet momenta: $dL(l;q,M)$.

We define the approximation appropriate to the region $R$ as a
replacement symbolized by an operation $T_R$:
\begin{eqnarray}
\label{eq:main.approx}
   \Gamma[W]
   ~\mapsto ~
   T_{R}\Gamma[W] &\equiv&  
            \int dL(\hat l;q,0)
            ~ H_R(\hat l, m\to 0) 
\nonumber\\
 && \hspace*{4mm}
            \times
            \prod_j \left[
                 \int \frac {dM_j^2}{2\pi }
                 \int dL(p_j;l_j,m) \Theta(M_j^2/\mu_F^2) 
                 J_{Rj}(p_j)
             \right]
             W(f).
\end{eqnarray}
This approximation is to be thought of as a kind of Taylor expansion
of the hard part $H_R$.  The hard subgraph and the associated
phase-space integrations have been replaced by massless
approximations, and $\hat l$ is used to denote the massless
approximated momenta that correspond to the incoming momenta of the
jets.  A precise definition of the relation between the approximated
and unapproximated momenta will be given below.  Note that the
unapproximated jet momenta still have their physical non-zero
invariant masses: $l_j^2=M_j^2$.

In a region where the final-state lines of the subgraphs $J_{Rj}$ do in fact
form jets, $T_R\Gamma$ provides a good approximation to the original graph
$\Gamma$, up to power-law corrections.  However, we will need to use
$T_R\Gamma$ outside the domain of validity of the approximation, and then
the integrals over the jet masses $M_j$ are no longer restricted by
momentum conservation.  It is therefore necessary to introduce a cut-off
function $\Theta(M_j^2/\mu_F^2)$ for large jet masses.  Here $\mu_F$ is
the so-called ``factorization scale'', which in principle is arbitrary.
The function $\Theta(x)$ must equal one for small values of $x$ and zero
for large values of $x$, and for this an ordinary $\theta$ function is
suitable\footnote{ Hence the notation!  }: $\Theta(x) = \theta(1-x)$.
However a smoother function would also be appropriate, and is likely to be
better for numerical work \cite{MC.BGF}.

The factorization scale determines the apportioning of the cross section
between LO and NLO (and higher) corrections.  As usual there is a
renormalization-group invariance of the complete cross section under
change of $\mu_F$: When all orders of perturbation theory are included, the
factorized cross section is accurate up to power-law corrections.  However
truncation of a perturbation expansion produces errors of the order of the
next higher power of $\alpha_s$, and an inappropriate choice of $\mu_F$ will
produce logarithmically large perturbative corrections.  For the hard
scattering, $\mu_F$ should therefore be around $Q$, or a bit smaller, if the
choice $\Theta(x) = \theta(1-x)$ is made.  This is because $Q$ sets the typical
scale for virtualities in a hard scattering.  A change in the functional
form of $\Theta$ simply corresponds to a more general renormalization-group
transformation.

It may be useful to apply a constraint on $\mu_F$ so that the momenta
automatically avoid violating the constraints given by momentum
conservation.  For this we choose $\mu_F$ so that $\Theta(M^2/\mu_F^2)$ is
zero whenever $M > Q/N_R$, where $N_R$ is the number of jet subgraphs for
the region $R$.

\subsection{Definition of projection onto massless momenta} 
\label{sec:projection}

The main element of the approximation $T_{R}\Gamma[W]$ in Eq.\ 
(\ref{eq:main.approx}) is the replacement of the external momenta of
$H_R$ by a suitable massless and on-shell approximation; let us notate 
this by
\begin{equation}
   \hat l = P_{N}(l) .
\end{equation}
The operator $P_{N}$ is a kind of (non-linear) projection, and the
definition\footnote{  
   It is likely that other definitions are possible and even useful.
   However, that issue will not be explored here.
}
that I choose is that in the overall center-of-mass frame
\begin{eqnarray}
\label{eq:l.hat}
    \hat{\JCCvec{l}}_j &=& \lambda^{-1}\JCCvec{l}_j ,
\nonumber\\
    \hat{l}_j^{0} &=& \left| \hat{\JCCvec{l}}_j \right | ,
\end{eqnarray}
with
\begin{equation}
\label{eq:lambda}
   \lambda  = \sum_j \left| \JCCvec{l}_j \right| / Q .
\end{equation}
That is, the spatial momenta are simply scaled by a common factor $\lambda$ so
that the total spatial momentum remains zero.  The scaling factor $\lambda$ is
chosen to give the correct total energy, so that the total energy-momentum
is preserved:
\begin{equation}
   \sum_j \hat{l}_j^\mu
   = \sum_j l_j^\mu
   = (Q, \JCCvec{0}) .
\end{equation}
The physical requirement of conservation of energy and momentum has the
mathematical consequence that the delta function 
$\delta^{(6)}\!(\sum_j l_j - q)$
in the phase-space integral remains unchanged, apart from a Jacobian
factor, after the jet momenta $l_j$ are replaced by their massless
approximations $\hat{l}_j$.  Note also that the operator $P_N$ satisfies
the relation $P_N(P_N(l)) = P_N(l)$ characteristic of a projection.

The derivation of the factorization theorem will rely on the fact that
the transformation $l \mapsto \hat l$ has a smooth limit when some or
any of the masses of the momenta go to zero.  This enables us to
use an expansion in powers of masses relative to $Q$ at appropriate
places. 

The inverse transformation from $\hat{l}$ to $l$ will be used in the
Monte-Carlo algorithm.  It is given by
\begin{eqnarray}
\label{eq:lj.construction}
    \JCCvec{l}_j &=& \lambda \hat{\JCCvec{l}}_j ,
\nonumber\\
    l_j^{0} &=& \sqrt {\JCCvec{l}_j^2 + M_j^2} ,
\end{eqnarray}
with $\lambda$ now being the solution of
\begin{equation}
\label{eq:lambda.def2}
   \sum_j \sqrt {{\strut \lambda}^2 {\strut\hat{\JCCvec{l}}_j}^2 
                    + M_j^2}
   = Q.
\end{equation}
Notice that this last equation has a solution if and only if 
$\sum_jM_{j} \leq Q$, and that then the solution is unique, with 
$0 \leq \lambda \leq 1$.

\subsection{Jacobian} 
\label{sec:jacobian}

As we will see in later sections, a contribution to a hard scattering
coefficient contains the value of an actual Feynman graph together
with subtractions that prevent double counting of lower-order
contributions.  Internal lines in the original graph will have their
exact energies and momenta, while the subtraction terms will have
approximated energies and momenta.  However, in a calculation of the
hard scattering coefficient we must use the same phase-space measure
in both the original graph and the subtractions.  Therefore we will
need the Jacobian of the transformation between the exact and the
approximated parton momenta:
\begin{equation}
\label{eq:phase.space.transformation}
  dL(\hat l;q,0) = dL(l;q,M) \Delta(l) ,
\end{equation}
so that, for example, Eq.\ (\ref{eq:main.approx}) can be written as
\begin{equation}
\label{eq:main.approx.renotated}
   \Gamma[W]
   ~\mapsto ~
   T_{R}\Gamma[W] \equiv
            \sum_f \Delta(l) 
            ~ H_R(\hat l, m=0)
            \prod_j \left[
                 \Theta(l_j^2/\mu_F^2) 
                 J_{Rj}(p_j)
             \right]
             W(f) ,
\end{equation}
where the sum over $f$ means an integral over all final-states with
the correct phase-space measure $dL(l;q,M)$, just as in Eq.\
(\ref{eq:sigma.W}). 

Some reasonably straightforward manipulations show that the Jacobian is
\begin{equation}
\label{eq:Jacobian.6.dim}
   \Delta(l) = \left( \prod_j \frac{ l_j^0 }{ |\JCCvec{l}_j| } \right)
         \sum_j \frac{ \JCCvec{l}_j^2}{ l_j^0 }
         \frac{ \lambda^{5-4N_R} }{ Q } .
\end{equation}
Here again $N_R$ is the number of jet subgraphs for the region in
question, and $\lambda$ is defined by Eq.\ (\ref{eq:lambda}).  In a general
number of space-time dimensions $D$, the formula remains the same,
except that the factor of $\lambda^{5-4N_R}$ is replaced by
\begin{equation}
\label{eq:Jacobian.factor.d.dim}
  \lambda^{D-1-(D-2)N_R}.
\end{equation}

In the massless limit, the Jacobian approaches unity:
\begin{equation}
   \Delta(l) = 1 + O(M^2/Q^2) ,
\end{equation}
where the correction term contains a term for each of the masses. 

In the case of a region with two jet subgraphs, $N_R=2$, these
formulae simplify, and we have
\begin{equation}
\label{eq:2jet.lambda}
   \lambda(\mbox{2 jets})
   ~=~
   \frac{2 |\JCCvec{l}_1|}{Q}
   ~=~
   \sqrt{ \left[ 1 - \frac{ (M_1+M_2)^2 }{ Q^2 } \right]
          \left[ 1 - \frac{ (M_1-M_2)^2 }{ Q^2 } \right]
        } ,
\end{equation}
and
\begin{equation}
\label{eq:2jet.Delta}
   \Delta(\mbox{2 jets})
   ~=~
   \left( \frac{ Q }{ 2 |\JCCvec{l}_1| } \right)^3
   ~=~
      \left[ 1 - \frac{ (M_1+M_2)^2 }{ Q^2 } \right]^{-3/2}
      \left[ 1 - \frac{ (M_1-M_2)^2 }{ Q^2 } \right]^{-3/2}
   .
\end{equation}

\subsection{Jet factors and factorization}
\label{sec:jet.factors}

With the above definitions, we can rewrite the factorization formula Eq.\ 
(\ref{eq:basic.factorization}) in terms of phase-space integrals for
massless partons and of factorized integrals for each jet.  
\begin{equation}
\label{eq:factorization2}
   \sigma[W] = K \sum_{N=2}^{\infty }
          \int dL(\hat l;q,0) 
          ~ \hat{H}_N
          \prod_j\left[
             \int \frac {dM_j^2}{2\pi }
             \int dL(p_j;l_j) \Theta(l_j^2/\mu_F^2) J_j(p_j)
         \right]
         ~ W(f).
\end{equation}
The use of the approximated phase-space measure, $dL(\hat l;q,0)$, for 
the external lines of the hard scattering is important, since it keeps 
the algorithm in the event generator simple: the generation of the
hard scattering part of the event can be performed without knowledge
of the subsequent showering and hadronization of the external
partons.  


%% file: phi3-examples.tex

\section{Examples}
\label{sec:examples}

In this section, I will show how the methods used in the proof of the
factorization theorem are applied to particular examples of Feynman
graphs.  This will provide detailed motivation for an appropriate
construction of the contribution for each particular region of momentum
space, and in particular for the form of the subtraction terms.

\subsection{Need for subtractions}
\label{sec:need.subtractions}

A initial attempt to obtain a factorization theorem goes as follows:
First, sum Eq.\ (\ref{eq:main.approx}) over all possible jet
configurations.  Then, for each configuration, sum over all possible
graphs compatible with the configuration.  As follows from the argument in
Sec.\ \ref{sec:proof.summary}, the result would appear to have the
factorized form given in Eq.\ (\ref{eq:basic.factorization}) or
(\ref{eq:factorization2}), provided that we define the contribution
associated with region $R$ to be
\begin{equation}
  \label{eq:naive.CR}
  C_R\Gamma(\mbox{region method}) 
  = T_R\Gamma \bigg|_{{\rm restricted~to~region~}R} .
\end{equation}
We would then define the hard scattering coefficient $\hat H_N$ to be
simply the sum over all graphs that make $N$ massless outgoing partons,
but with a restriction on the momenta to be in the appropriate hard
region.  There is a corresponding restriction to collinear configurations
in the jet factor.

Although this simple phase-space-slicing argument contains a core of
truth, it fails to be exactly correct, for several interrelated reasons:
\begin{itemize}

\item 
  The boundaries of the various regions are hard to define precisely for
  the general case, and most importantly the regions are liable not to
  factorize into hard and collinear parts.

\item  
  If one does require a factorized form for the shapes of the regions in
  momentum space, it is possible that there can be double counting of some
  parts of momentum space, and omissions of other parts.  It is quite
  unobvious how to automatically avoid such double- and/or non-counting in
  general, with a phase-space-slicing method.  
   
\item 
  A particular jet subgraph may correctly lie just below a threshold
  defining a jet, but some lines in the hard part may have invariant
  masses only just above the threshold.  Then it is far from clear that
  the massless approximation for the external lines of the hard part
  results in a power-suppressed error.  So there is a danger that the
  factorization theorem proved with the phase-space-slicing method fails
  to accomplish its purpose of providing an approximation to the cross
  section that is valid to the whole leading power.

\end{itemize}

Instead, we will use a subtractive method.  We will still construct the
asymptotic form of the cross section as a sum over terms, one for each
region of each graph, as in Eqs.\ (\ref{eq:sum.over.regions}) and
(\ref{eq:basis.of.proof}).  Each term $C_R(\Gamma)$ will be exactly a
contribution to the hard scattering times a contribution to the jet
factors, and it will also equal the basic approximation $T_R\Gamma$ plus
subtraction terms that cancel the effects of double counting between
regions.  However, the use of a subtraction method allows the integrals
over momentum to be unrestricted.  The sum over all graphs and regions
then gives the factorization formula, Eq.\ (\ref{eq:basic.factorization}).
The combinatoric argument is the same as in the naive phase-space slicing
method, but now there is no restriction on the integrations, other than
from the cut-off function $\Theta$ in the definition of the jet factor---see
Eq.\ (\ref{eq:factorization2}) and Eq.\ (\ref{eq:jet.factor.def}) below.

\subsection{Simple example}
\label{sec:simple.example}

In this section, I will implement the subtraction method for the
simplest example, the cut graph, $A$, of Fig.\ \ref{fig:3jet}, whose
final state has three particles, of momenta $p_1$, $p_2$ and $p_3$.  I
will give a lot of explicit detail for what is actually a fairly
simple problem.  The reason is that the techniques will generalize to
much more complicated situations, but only a precise treatment will
make this generalization clear.

The graph has the following regions for its leading twist
contributions:
\begin{enumerate}

\item[$R_1$:]
    $p_1$ and $p_2$ are almost parallel.

\item[$R_2$:]
    All the final-state momenta are at wide angles with respect to
    each other.  This is a purely ultra-violet region where the graph
    can be correctly approximated by setting all masses to zero.

\end{enumerate}

We will see how, in the leading power of $Q$, $\Gamma$ can be written
as a sum of a term for each of the two regions, plus a suppressed
remainder term:
\begin{equation}
\label{eq:A.decomp}
   A = A_1 + A_2 
   + O\!\left( \left| A \right| \frac{ m^2 }{ Q^2 } \right) .
\end{equation}
Here, each of $A_1$ and $A_2$ is a particular term on the right-hand side
of Eq.\ (\ref{eq:sum.over.regions}), with $A_j$ meaning $C_{R_j}(A)$.

\begin{figure}
\psfrag{k_a}{$k_a$}
\psfrag{k_b}{$k_b$}
\psfrag{p_1}{$p_1$}
\psfrag{p_2}{$p_2$}
\psfrag{p_3}{$p_3$}
\includegraphics{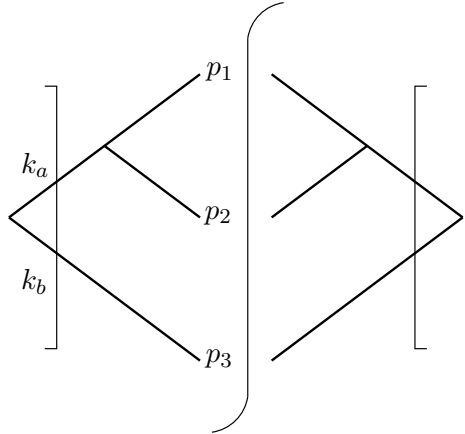}
\caption{First leading region of Fig.\ \protect\ref{fig:3jet}, and its
         standard approximation. The vertical bars with the squared
         ends indicate where the collinear approximation of Eq.\
         (\ref{eq:main.approx}) is made.}
\label{fig:A.1}
\end{figure}

\paragraph*{Region $R_1$}
In this region, a good approximation, symbolized in Fig.\
\ref{fig:A.1}, is
\begin{equation}
\label{eq:A.1}
   A_1
   \equiv T_{R_1} \Gamma
   = dL(k_a,k_b;q,0)
     ~ H_{A1}(k_a, k_b)
     ~ dD_{12}(p_1,p_2) ,
\end{equation}
where I have used the definition, Eq.\ (\ref{eq:main.approx}), for the
Taylor expansion operator, $T_{R_1}$.

In this approximation the two external momenta of the hard subgraph are
replaced by on-shell massless momenta:
\begin{eqnarray}
   k_a^\mu &~=~&
   \frac { \Big(|\JCCvec{p}_1 + \JCCvec{p}_2|,~ 
                \JCCvec{p}_1 + \JCCvec{p}_2 
           \Big)    
   }{ \lambda_{A1}} 
   ~=~ \frac {Q}{2}
       \Big(
            1,~
            - \JCCvec{p}_3/ |\JCCvec{p}_3|
       \Big),
\\
    k_b^\mu &~=~& \frac {Q}{2}
              \Big(
               1,~
               \JCCvec{p}_3/ |\JCCvec{p}_3|
              \Big) .
\end{eqnarray}
The scaling parameter $\lambda_{A1}$ that appears in the transformation
between the exact jet momenta and the massless approximation has the
value
\begin{equation}
\label{eq:A.lambda.1}
    \lambda_{A1} = \frac {2 |\JCCvec{p}_3|}{Q} .
\end{equation}
In the projection notation, 
$(k_a, k_b) = P_2(p_1+p_2, \, p_3)$. 

The hard scattering factor in Eq.\ (\ref{eq:A.1}) is simply unity, which
is the value of the lowest order graph.  The jet factor is defined by a
suitable specialization of the factor in square brackets in Eq.\ 
(\ref{eq:factorization2}):
\begin{equation}
\label{eq:D12.def}
   dD_{12} = \int \frac {dM_{12}^2}{2\pi }
               ~dL(p_1, p_2; l_{12}) 
               ~\Theta(M_{12}^2/\mu_F^2) 
               ~ \frac{g^2}{[(p_1+p_2)^2-m^2]^2} ~.
\end{equation}
Here, $l_{12}^\mu \equiv p_1^\mu + p_2^\mu$ is the exact momentum of the
intermediate parton.  The errors in the approximation are solely in the
massless approximation to the two-particle phase space, so that we can
write
\begin{equation}
    A = A_1 \left[ 1 + O(M_{12}^2/Q^2) \right] .
\end{equation}
Here $M_{12}$ is the invariant mass of the final state in the jet
subgraph, i.e., $\sqrt{(p_1+p_2)^2}$.

When we construct the subtractions for the second region, we will need
the formula for the approximation in terms of the original phase-space
integration:
\begin{equation}
    A_1 = dL(p;q,m)
           ~ \Delta_{A1}
           ~ \Theta((p_1+p_2)^2 / \mu_F^2 )
           ~ \frac{ g^2 } { [(p_1+p_2)^2-m^2]^2 } ~,
\end{equation}
where the Jacobian is
\begin{equation}
   \Delta_{A1} = \left( \frac{ Q }{ 2 | \JCCvec{p}_3 |}
                 \right)^3 ,
\end{equation}
from Eq.\ (\ref{eq:2jet.Delta}).

\paragraph*{Region $R_2$}
In this region, all the external particles are at wide angle, so that
the appropriate approximation to the whole graph is
\begin{equation}
\label{eq:T.A.2}
    T_{R_2} A
   = dL(l_{2,a}, l_{2,b}, l_{2,c}; q, 0)
   ~ H_{A2}(l_{2,a}, l_{2,b}, l_{2,c}) .
\end{equation}
The hard subgraph $H_{A2}$ is the whole graph, taken in the massless
approximation: 
\begin{equation}
   H_{A2} = \frac{g^2}{ \left[ (l_{2,a} + l_{2,b})^2 \right]^2 } ~,
\end{equation}
and its external momenta are:
\begin{eqnarray}
     l_{2,a}^\mu &=&
     \frac { \Big( |\JCCvec{p}_1|,~ \JCCvec{p}_1 \Big) }{ \lambda_{A2}} ~,
\\
     l_{2,b}^\mu &=&
     \frac { \Big( |\JCCvec{p}_2|,~ \JCCvec{p}_2 \Big) }{ \lambda_{A2}} ~,
\\
     l_{2,c}^\mu &=&
     \frac { \Big( |\JCCvec{p}_3|,~ \JCCvec{p}_3 \Big) }{ \lambda_{A2}} ~,
\end{eqnarray}
where
\begin{equation}
    \lambda_{A2}
    = \frac { |\JCCvec{p}_1| + |\JCCvec{p}_2| + |\JCCvec{p}_3| }
          {Q} 
    = 1 + O(m^2/Q^2) .
\end{equation}

Now we note two related facts
\begin{itemize}

\item $H_{A2}$ has a collinear singularity when $l_{2,a}$ and $l_{2,b}$
   become parallel.

\item Term $A_1$, which I have constructed to be a good approximation
   in region $R_1$, also contributes in the larger region $R_2$.

\end{itemize}

So I define $A_2$ by approximating the original graph minus the
approximation I have already constructed for the smaller region
$R_1$.  The simplest definition would be
$A_2 = T_{R_2} \! \left( A - T_{R_1} A \right)$.  However a small
modification turns out to be convenient, and I define
\begin{eqnarray}
\label{eq:A.2}
   A_2 &=& 
      V(\widetilde M_{12} / m)
      ~ T_{R_2} \! \left( A - T_{R_1} A \right) 
\nonumber\\
  &=& dL(p;q) 
      ~ V(\widetilde M_{12} / m)
      ~ \Delta_{A2}
      ~ \frac{ g^2 }
             { \left[ (l_{2,a} + l_{2,b})^2 \right]^2 }
        ~\left[
             1 - \Theta((l_{2,a}^2+l_{2,b})^2 / \mu_F^2) 
                 ~\Delta_{A1'}
        \right] .
\end{eqnarray}
Here $\Delta_{A2}$ is the Jacobian for the transformation $(p_1, p_2,
p_3) \mapsto (l_{2,a}, l_{2,b}, l_{2,c})$, which is equal to unity
plus corrections of order $m^2/Q^2$.  The Jacobian $\Delta_{A1'}$ in
the subtraction term is the approximation to $\Delta_{A1}$ in which the
momenta $p_1$, etc., are replaced by their massless approximations
$l_{2,a}$, etc., i.e.,
\begin{equation}
   \Delta_{A1'} = \left( \frac{ Q }{ 2 | \JCCvec{l}_{2,c} | }
              \right)^3 
          = \Delta_{A1} \times \left[ 1 + O(m^2/Q^2) \right],
\end{equation}
The remaining factor $V(\widetilde M_{12}, m)$ is what I will call a
``collinear veto factor''.  Its purpose is to remove a spurious collinear
divergence.  After explaining the need for this factor, I will show that
the sum of $A_1$ and $A_2$ provides a good approximation everywhere, Eq.\ 
(\ref{eq:A.decomp}).

First, recall that $A_1$ is obtained from $A$ by the approximation of
neglecting both the jet mass $M_{12}$ and the particle masses with respect
to $Q$.  So the sources of error give terms of order $M_{12}^2/Q^2$ or
$m^2/Q^2$.  Since $M_{12}$ is always larger than $2m$, we can write
\begin{eqnarray}
\label{eq:A.minus.A1}
  A-A_1 &=&
  O\!\left( \left|A\right| \frac{ M_{12}^2 }{ Q^2 } \right) .
\end{eqnarray}
We need to assume here that the factorization scale $\mu_F$ is of order
$Q$; otherwise we would have a large logarithmic contribution to the
total cross section.

Next, we observe that $A_2$ is obtained from $A-A_1$ by applying a
massless approximation for the external particles, as is appropriate when
all the particles are at wide angles.  This approximation results in an
approximated jet mass $\widetilde M_{12}$ that is different from $M_{12}$.
When the final-state particles are at wide angles, the exact and
approximated masses are equal up to corrections of relative size
$m^2/Q^2$.  But as the two particles become exactly parallel, the
approximated jet mass goes to zero and produces a singularity in the
propagator denominator, whereas the exact jet mass never goes below $2m$.
Thus, without the veto factor, $A_2$ has a singularity in the collinear
region.  Indeed, when we replace the correct propagators 
$1 / \left[ (p_1+p_2)^2-m^2 \right]^2$ in the original graph $A$ by their
massless approximation $1/\left[ (l_{2,a}+l_{2,b})^2 \right]^2$, we get a
well-known logarithmic collinear divergence in the integral over all jet
masses.

However, the estimate Eq.\ (\ref{eq:A.minus.A1}) ensures that there is a
power-law suppression:
\begin{eqnarray}
\label{eq:A2.estimate}
    A_2 
    &=& O\!\left( \left| A-A_1 \right| 
                  \frac{ m^4 }{ \widetilde M_{12}^4 } 
           \right) 
\nonumber\\
    &=& O\!\left( \left| A \right| 
                 \frac{ \widetilde M_{12}^2 }{ Q^2 }
                 \frac{ m^4 }{ \widetilde M_{12}^4 } 
         \right) .
\nonumber\\
    &=& O\!\left( \left| A(m=0) \right| 
                 \frac{ \widetilde M_{12}^2 }{ Q^2 }
         \right) .
\end{eqnarray}
The estimate on the first line results from the replacement of the exact
propagator by the massless approximation, and is valid when $\widetilde
M_{12} \lesssim m$.  The second line results from applying Eq.\
(\ref{eq:A.minus.A1}) in the massless approximation.  

Although we now see that the logarithmic divergence in the integrated
cross section has been removed, there remains a divergence in the
differential cross section.  This can be quite annoying in practice, and
it also complicates the proofs and error estimates.  So we simply remove
the divergence by defining $A_2$ to include a veto factor $V(\widetilde
M_{12} / m)$.  One suitable veto factor is simply a theta function
\begin{equation}
\label{eq:veto.function}
  V(\widetilde M_{12} / m) = \theta( \widetilde M_{12} - 2m ).  
\end{equation}
Any other function satisfying the following properties would be equally
good:
\begin{itemize}
\item $V$ is zero when $\widetilde{M}_{12}$ is significantly less than
  $m$.
\item $V$ is unity when $\widetilde{M}_{12}$ is significantly greater than
  $m$.
\end{itemize}
The first requirement ensures that the divergences are removed.  Indeed
the whole region when $A_2$ is much larger than $A-A_1$ is also removed,
i.e., the region where a massless approximation to $A-A_1$ is a disaster.
Since the purpose of $A_2$ is to handle momentum configurations far from
the collinear region, the precise functional form of $V$ in the collinear
region is irrelevant to physics.  The second requirement, that $V$ be
unity away from the collinear region, ensures that $A_2$ accomplishes its
purpose of participating in a better approximation to $A$ in the
non-collinear region, $R_2$.  

The veto factor ensures that the approximated jet mass $\widetilde M_{12}$
is never much smaller than $m$, i.e., $\widetilde M_{12} \gtrsim m$.  In this
case, the true and approximated jet masses are comparable.  Thus, when we
set $m$ to zero in calculating $A_2$, we are neglecting $m$ in comparison
with the smallest virtuality, which is always of order $M_{12}$.  Hence
\begin{equation}
    A_2 - (A-A_1) 
    = O\!\left( \frac{ m^2 }{ M_{12}^2 } \left|A-A_1\right| \right)
    = O\!\left( \frac{ m^2 }{ Q^2 } \left|A\right| \right) ,
\end{equation}
where we have used Eq.\ (\ref{eq:A.minus.A1}).  This is exactly the
desired result, Eq.\ (\ref{eq:A.decomp}).  

Notice how the exact form of the veto factor does not affect this
derivation.  A change in the veto factor only affects the collinear
region, $M_{12} \sim m$.  Although the change might result in a 100\%
change in $A_2$, this does not affect the error estimate, since both
$A-A_1$ and $A_2$ are inevitably of order $m^2/Q^2$ (relative to $A$)
in this region.

\subsection{Harder example}
\label{sec:harder.example}

In this section, we treat a more complicated example, the cut graph,
$\Gamma$, of Fig.\ \ref{fig:4jet}, which has four particles in the final
state, of momenta $p_1$, $p_2$, $p_3$ and $p_4$.  As was observed
earlier, the graph has the following regions for leading twist
contributions:
\begin{enumerate}

\item[$R_1$:]
    $p_1$ and $p_2$ are almost parallel, and $p_3$ and $p_4$
    are also almost parallel.

\item[$R_2$:]
    $p_3$ and $p_4$ are almost parallel, but $p_1$ and $p_2$
    are at wide angle.

\item[$R_3$:]
    $p_1$ and $p_2$ are almost parallel, but $p_3$ and $p_4$ are at
    wide angle.

\item[$R_4$:]
    All the final-state momenta are at wide angles with respect to
    each other.  

\end{enumerate}
Our treatment will start off being almost identical to that for Fig.\
\ref{fig:3jet}, in Sec.\ \ref{sec:simple.example}.  But complications will
arise later because there is the equivalent of an overlapping divergence. 

\begin{figure}
   \psfrag{R1}{$R_1$}
   \psfrag{R2}{$R_2$}
   \psfrag{R3}{$R_3$}
   \psfrag{R4}{$R_4$}
   \includegraphics[width=1.3in]{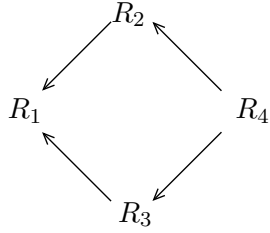}
   \caption{Relation between leading regions of Fig.\
            \protect\ref{fig:4jet}.}
   \label{fig:relation}
\end{figure}

\begin{figure}
\includegraphics{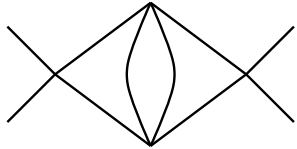}
\caption{Graph in $\phi^4$ theory with overlapping ultra-violet
   divergences.}
\label{fig:overlapping.divergences}
\end{figure}

As before, I will show how, in the leading power of $Q$, $\Gamma$ can be
written as the sum of a term for each of the four regions, plus a
suppressed remainder term:
\begin{equation}
\label{eq:Gamma.decomp}
   \Gamma = \Gamma_1 + \Gamma_2 + \Gamma_3 + \Gamma_4
            + O\!\left( \left| \Gamma \right| \frac{ m^2 }{ Q^2 } \right).
\end{equation}
In the more general notation of Eq.\ (\ref{eq:sum.over.regions}), we would
have $C_{R_1}\Gamma$ instead of $\Gamma_1$, etc.  To organize the
subtractions in these terms, we need an ordering relation between the
regions, which is essentially set theoretic inclusion.  For the graph
under discussion, this relation is symbolized in Fig.\ \ref{fig:relation},
and I notate the corresponding orderings as $R_1<R_2$, $R_1<R_3$,
$R_2<R_4$ and $R_3<R_4$.  This means that $R_1$ is contained in both $R_2$
and $R_3$, and that both $R_2$ and $R_3$ are contained in $R_4$.  This
same structure appears in simple examples of overlapping ultra-violet
divergences.  (For example it gives the relationships between the regions
of UV divergence for the graph of Fig.\
\ref{fig:overlapping.divergences}.)  Thus the example of Fig.\
\ref{fig:4jet} provides an interesting and non-trivial test of our
methods.

A general and precise definition of the ordering is made by associating
each region with a particular set of momentum configurations of massless
momenta, and then defining the ordering as simple set theoretic inclusion
of these massless configurations.  The configurations associated with a
region $R$ are those where the momenta within each jet are made massless
and exactly parallel.  These are exactly the configurations that are used
to classify the pinch singularities of massless graphs in the
Libby-Sterman analysis \cite{pc}. For example, the configurations for
$R_1$ are all those for which $p_1$ and $p_2$ are exactly parallel and
massless and $p_3$ and $p_4$ are exactly parallel and massless.
Similarly, the configurations for $R_2$ are those for which $p_3$ and
$p_4$ are exactly parallel.  Thus one set of momenta is contained in the
other: $R_1 \subset R_2$, and I use this as the definition\footnote{
   Note carefully the distinction between the massless collinear
   configurations that are used to define relations like $R_1 < R_2$
   and the corresponding sets of momenta that give the 
   important regions contributing to the cross section.  The massless
   collinear configurations form skeletons, so to speak, for the
   regions.  Defining the relations $R_1 < R_2$ by set-theoretic
   inclusion for the skeletons is exactly literally correct.  But the same
   definition applied to the regions requires careful interpretation to
   deal with the boundaries etc.  (The concept of a massless collinear
   momentum configuration is precisely defined, but the concept of an
   important momentum region is rather fuzzy and not susceptible to a
   precise and useful mathematical definition.) 
   \\
   A related issue is the commonly seen misunderstanding that the
   factorization theorem is concerned with the contributions of collinear
   (and soft) divergences.  Literally speaking, this assertion is false,
   as is clear in any field theory that has non-zero masses for all its
   fields: these preclude there from being actual collinear and soft
   divergences, even though the factorization theorem remains valid.  It
   is true that the leading contributions are {\em associated} with
   divergences.  But the divergences are not necessarily literally present
   in the actual theory, and a correct treatment cannot assume this.  The
   correct statement is that leading regions of momentum space can be
   labeled by the skeleton configurations that are surfaces of pinch
   singularities in massless Feynman graphs, and that the leading regions
   are contained in neighborhoods of the skeleton configurations.
} of $R_1 < R_2$.

\begin{figure}
   \includegraphics[width=7cm]{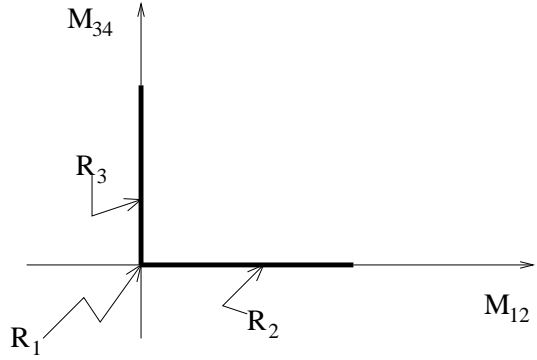}
   \caption{Geometry of the leading regions for Fig.\
     \protect\ref{fig:4jet}. }
   \label{fig:region.geometry}
\end{figure}

In the case of the regions for our graph, a convenient visualization of
the geometry is given in Fig.\ \ref{fig:region.geometry}.  This shows the
plane of the two relevant variables, the mass $M_{12}$ of the $(12)$ pair
and the mass $M_{34}$ of the $(34)$ pair.  The singular surface associated
with $R_1$ is the point $M_{12} = M_{34} = 0$; the surface associated with
$R_2$ is the line $M_{34} = 0$, restricted to the physically accessible
region; the surface associated with $R_3$ is the line $M_{12} = 0$, again
restricted to the physically accessible region; and the ultra-violet
region $R_4$ is the whole plane.  Set theoretic relations between these
sets of points evidently give $R_1<R_2$, $R_1<R_3$, $R_2<R_4$ and $R_3<R_4$.

The construction of Eq.\ (\ref{eq:Gamma.decomp}) will use this ordering.
I will start with the smallest region, $R_1$, and construct its term $\Gamma_1$
to be a good approximation to $\Gamma$ in this region.  Then I will
successively construct the terms for larger regions.  Each term will be
arranged so that, when added to the previously constructed terms, it does
not violate the validity of the approximations provided by these terms.
The first part of the treatment is therefore very close to that of the
lower-order graph Fig.\ \ref{fig:3jet}.

\begin{figure}
\psfrag{k_a}{$k_a$}
\psfrag{k_b}{$k_b$}
\psfrag{p_1}{$p_1$}
\psfrag{p_2}{$p_2$}
\psfrag{p_3}{$p_3$}
\psfrag{p_4}{$p_4$}
\includegraphics{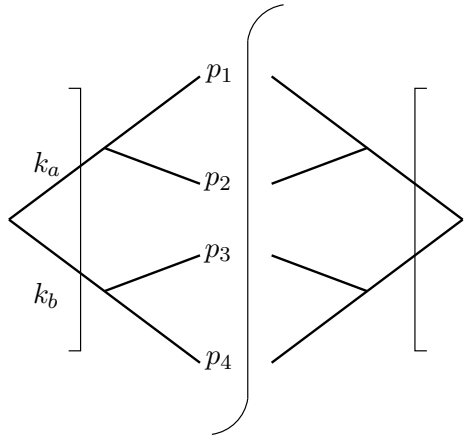}
\caption{First leading region of Fig.\ \protect\ref{fig:4jet}, and its
         standard approximation.}
\label{fig:region.1}
\end{figure}

From the Feynman rules of the theory, we see that the contribution of
the unapproximated graph to the cross section is (up to overall
kinematic factors):
\begin{equation}
   \Gamma
=  dL(p;q,m)
   ~ \frac {g^{4}}
           {  \left[ (p_1+p_2)^2 - m^2\right]^2
              ~ \left[ (p_3+p_4)^2 - m^2\right]^2
           } ~.
\end{equation}

\paragraph*{Region $R_1$}
In this region, a good approximation, symbolized in Fig.\
\ref{fig:region.1}, is
\begin{equation}
\label{eq:approx.1}
   \Gamma_1
   \equiv T_{R_1} \Gamma
   = dL(k_a,k_b;q,0)
     ~ H_1(k_a, k_b)
     ~ dD_{12}(p_1,p_2)
     ~ dD_{34}(p_3,p_4) .
\end{equation}
It is the obvious generalization of $A_1$, in Eq.\ (\ref{eq:A.1}), and
it is again constructed by replacing the external momenta of the hard
subgraph by on-shell massless momenta, which are now
\begin{eqnarray}
   k_a^\mu &=&
   \lambda_1^{-1}
   \Big(|\JCCvec{p}_1 + \JCCvec{p}_2|,
        ~ \JCCvec{p}_1 + \JCCvec{p}_2 
   \Big)
\nonumber\\
            &=& \frac {Q}{2}
              \left(
               1,~
               \frac{ \JCCvec{p}_1 + \JCCvec{p}_2 }
                    { \left|\JCCvec{p}_1 + \JCCvec{p}_2 \right| }
              \right),
\\
    k_b^\mu &=&
   \lambda_1^{-1}
   \Big(|\JCCvec{p}_3 + \JCCvec{p}_4|,
        ~ \JCCvec{p}_3 + \JCCvec{p}_4
   \Big) 
\nonumber\\
            &=& \frac {Q}{2}
              \left(
               1,~
               - \frac{ \JCCvec{p}_1 + \JCCvec{p}_2 }
                      { \left| \JCCvec{p}_2 + \JCCvec{p}_2 \right| }
              \right) .
\end{eqnarray}
Here
\begin{eqnarray}
    \lambda_1 &=& \frac {2 |\JCCvec{p}_1 + \JCCvec{p}_2|}{Q} ~.
\end{eqnarray}
In the projection notation, 
$(k_{a}, k_{b}) = P_2(p_1+p_2, \, p_3+p_4)$.

The hard scattering factor $H_1$ in Eq.\ (\ref{eq:approx.1}) is simply
unity, which is the value of the lowest order graph.  The jet factors are
defined by Eq.\ (\ref{eq:D12.def}) for $dD_{12}$, and a corresponding
formula for $dD_{34}$.  The errors in the approximation are solely in the
massless approximation to the phase space of the hard scattering.
Therefore we have
\begin{equation}
    \Gamma = \Gamma_1 \left[ 1 + O(M_1^2/Q^2) \right] ,
\end{equation}
where $M_1$ is the larger of the masses of the two pairs of particles:
$M_1^2 = \max( M_{12}^2, M_{34}^2 )$.

In constructing subtractions for other regions, we will need the
formula for the approximation in terms of the original phase-space
integration: 
\begin{equation}
    \Gamma_1 = dL(p;q,m)
           ~ \Delta_1
           ~ \frac{ g^2 \Theta( (p_1+p_2)^2/\mu_F^2 ) }
                  { [(p_1+p_2)^2-m^2]^2 }
           ~ \frac{ g^2 \Theta( (p_3+p_4)^2/\mu_F^2 ) }
                  { [(p_3+p_4)^2-m^2]^2 } ~,
\end{equation}
where the Jacobian is 
\begin{equation}
\label{eq:Delta1}
   \Delta_1 
   = \left( \frac{ Q }{ 2 |\JCCvec{p}_1 + \JCCvec{p}_2| } \right)^3 ,
\end{equation}
from Eq.\ (\ref{eq:2jet.Delta}).  

Observe that $\Gamma_1$ differs from the original graph only by the
Jacobian of the phase space and by the cut-offs for the jet factors.
So in terms of methods for calculating individual low-order Feynman
graphs, we have gained nothing; indeed we have lost something.  What
we have gained is a form that can be generalized to higher-order graphs
and converted to a systematic algorithm for obtaining useful
approximations to large numbers of graphs with a computational effort
merely proportional to the number of final-state particles.

\begin{figure}
\psfrag{k_{2a}}{$k_{2a}$}
\psfrag{k_{2b}}{$k_{2b}$}
\psfrag{k_{2c}}{$k_{2c}$}
\psfrag{p_1}{$p_1$}
\psfrag{p_2}{$p_2$}
\psfrag{p_3}{$p_3$}
\psfrag{p_4}{$p_4$}
\includegraphics{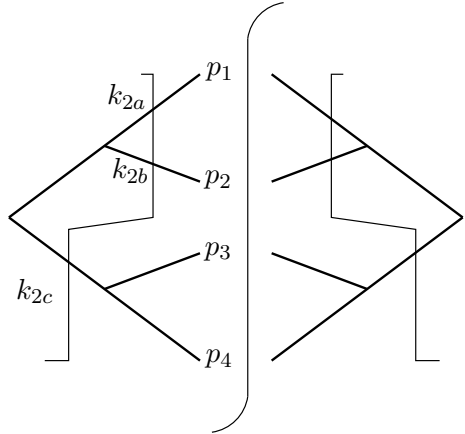}
\caption{Second leading region of Fig.\ \protect\ref{fig:4jet} and its
         standard approximation.}
\label{fig:region.2}
\end{figure}

\paragraph*{Region $R_2$}
In this region, where only particles 3 and 4 are collinear, the
appropriate approximation to the whole graph is symbolized in Fig.\
\ref{fig:region.2}:
\begin{equation}
\label{eq:approx.2}
    T_{R_2} \Gamma
   = dL(k_{2,a}, k_{2,b}, k_{2,c}; q, 0)
   ~ H_2(k_{2,a}, k_{2,b}, k_{2,c})
   ~ dD_{34}(p_3,p_4).
\end{equation}
The hard part $H_2$ is (the square of) the left-hand subgraph of Fig.\ 
\ref{fig:region.2}:
\begin{equation}
   H_2 = \frac{g^2}{ \left[ (k_{2,a} + k_{2,b})^2 \right]^2 } ~,
\end{equation}
whose external momenta are on-shell and massless:
\begin{eqnarray}
     k_{2,a}^\mu &=&
     \lambda_2^{-1}
     \Big( |\JCCvec{p}_1|,~ \JCCvec{p}_1 \Big) ,
\\
     k_{2,b}^\mu &=&
     \lambda_2^{-1}
     \Big( |\JCCvec{p}_2|,~ \JCCvec{p}_2 \Big) ,
\\
     k_{2,c}^\mu &=&
     \lambda_2^{-1}
     \Big( |\JCCvec{p}_3 + \JCCvec{p}_4|,
          ~  \JCCvec{p}_3 + \JCCvec{p}_4
     \Big)
,
\end{eqnarray}
where
\begin{eqnarray}
    \lambda_2 &=&
    \frac {|\JCCvec{p}_1| + |\JCCvec{p}_2| 
           + |\JCCvec{p}_3 + \JCCvec{p}_4|     }
          { Q }
`.
\end{eqnarray}

\begin{figure}
\begin{displaymath}
\psfrag{k_{2a}}{$k_{2a}$}
\psfrag{k_{2b}}{$k_{2b}$}
\psfrag{k_{2c}}{$k_{2c}$}
\psfrag{p_1}{$p_1$}
\psfrag{p_2}{$p_2$}
\psfrag{p_3}{$p_3$}
\psfrag{p_4}{$p_4$}
  \raisebox{-0.5\height}{\includegraphics{4jet-2.eps}} 
  ~~ - ~~
\psfrag{p_1}{$p_1$}
\psfrag{p_2}{$p_2$}
\psfrag{p_3}{$p_3$}
\psfrag{p_4}{$p_4$}
  \raisebox{-0.5\height}{\includegraphics{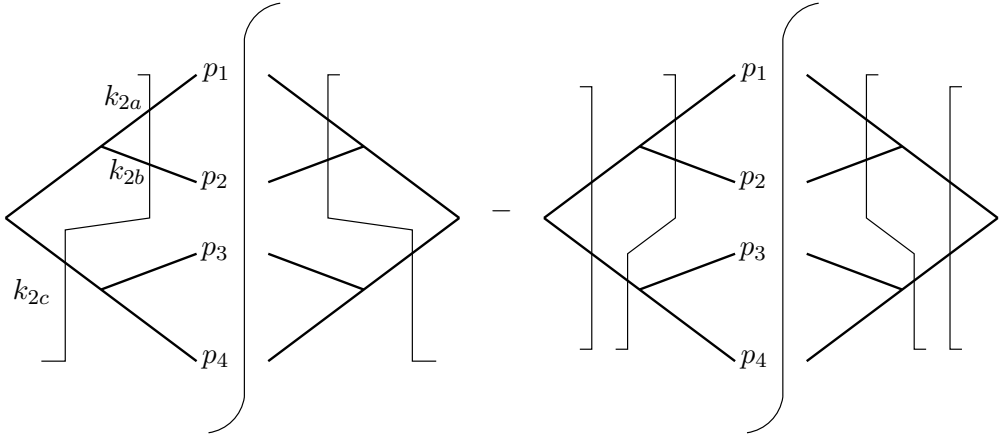}} 
\end{displaymath}
\caption{Definition of $\Gamma_2$.}
\label{fig:Gamma.2}
\end{figure}

Just as before, I define the contribution associated with region
$R_2$ by subtracting an allowance for the smaller region $R_1$:
\begin{eqnarray}
\label{eq:Gamma.2}
   \Gamma_2 &=& V_2 ~ T_{R_2} \!\left( \Gamma - \Gamma_1 \right)
\nonumber\\
   &=& V_2 ~ T_{R_2} \!\left( \Gamma - T_{R_1} \Gamma \right)
\nonumber\\
  &=& dL(p;q) 
      \Delta_2
      V_2
      \frac{ g^2 \Theta((p_3^2+p_4)^2 / \mu_F^2) }
             { [(p_3+p_4)^2-m^2]^2 }
      ~ \frac{ g^2 }
             { \left[ (k_{2,a} + k_{2,b})^2 \right]^2 }
      \times
\nonumber\\
   && \hspace{2cm} \times
      \left[
            1 - \Theta((k_{2,a}^2+k_{2,b})^2 / \mu_F^2) 
               ~\Delta_{1'}
      \right]
.
\end{eqnarray}
Here 
\begin{itemize}

\item The Jacobian for the transformation to $k_{2,a}$, $k_{2,b}$, and
   $k_{2,c}$ is
   \begin{equation}
   \label{eq:Delta2}
     \Delta_2 = 
     \frac{ E_1 }{ |\JCCvec{p}_1 | }
     \frac{ E_2 }{ |\JCCvec{p}_2 | }
     \frac{ E_3+E_4 }{ | \JCCvec{p}_3+\JCCvec{p}_4 | }
     \left(   
          \frac{ \JCCvec{p}_1^2 }{ E_1 }
          + \frac{ \JCCvec{p}_2^2 }{ E_2 }
          + \frac{ ( \JCCvec{p}_3+\JCCvec{p}_4 )^2 } { E_3+E_4 }
     \right)
     \frac{ 1 }{ \lambda_2^7 Q }
     ~,
   \end{equation}
   from Eq.\ (\ref{eq:Jacobian.6.dim}).

\item
   $\Delta_{1'}$ is $\Delta_1$ with the momenta $\JCCvec{p}_1+\JCCvec{p}_2$ and
   $\JCCvec{p}_3+\JCCvec{p}_4$ replaced by
   $\JCCvec{k}_{2,a}+\JCCvec{k}_{2,b}$ and $\JCCvec{k}_{2,c}$, i.e.,
   \begin{equation}
   \label{eq:Delta.1p}
     \Delta_{1'} = \left( \frac{ Q }{ 2 |\JCCvec{k}_{2,c}| } \right)^3 .
   \end{equation}

\item The veto factor $V_2$ is defined to be
   \begin{equation}
   \label{eq:V2}
      V_2 = V(\widetilde M_{12} / m ),
   \end{equation}
   where $V$ is the same veto function as in Eq.\ (\ref{eq:veto.function}).
   The reason for requiring a veto factor is the same as with Eq.\
   (\ref{eq:A.2}). 

\end{itemize}
The formula for $\Gamma_2$ is symbolized in Fig.\ \ref{fig:Gamma.2}.
It is power suppressed in the smaller region $R_1$, whereas the
massless approximations to the individual terms $\Gamma$ and
$\Gamma_1$ both have collinear singularities on the surface $R_1$.
The sum of the two terms I have generated so far gives a good
approximation to the original graph in a neighborhood of $R_2$ (and hence
of the smaller surface $R_1$, whose defining surface is a subsurface of
$R_2$): 
\begin{equation}
\label{eq:R2.decomp}
   \Gamma = (\Gamma_1 + \Gamma_2) 
       \left[ 1 + O(M_{34}^2/Q^2) \right],
\end{equation}
where $M_{34}$ is the mass of the pair of particles 3 and 4.  The proof of
this equation is the same as for the corresponding equation Eq.\ 
(\ref{eq:A.decomp}) for the simpler graph.  

A quick examination of Fig.\ \ref{fig:region.geometry} might suggest
that the there is some kind of singularity on the line $R_3$, and that
this would cause a problem in applying the above argument around the
point $R_1$.  This is not so.  The treatment of the double-collinear
region associated with $R_1$ applies uniformly in a whole neighborhood
of the point $R_1$.  The line $R_3$ gains its significance only when
one moves substantially away from $R_1$ and finds a region where only
$p_1$ and $p_2$ are collinear while the other momenta are at wide
angles.

There is an important transitivity property of successive projections
of momenta.  We obtained $k_a$ and $k_b$ by a projection of the four
momenta $p_1$, $p_2$, $p_3$ and $p_4$.  Instead we could project the
three momenta $k_{2,a}$, $k_{2,b}$, and $k_{2,c}$ onto two momenta by
\begin{eqnarray}
   k_{a'}^\mu
   &=&
   \lambda_{21}^{-1}
   \Big( \left| \JCCvec{k}_{2a} + \JCCvec{k}_{2b} \right|,
        ~ \JCCvec{k}_{2a} + \JCCvec{k}_{2b}
   \Big) ,
\\
   k_{b'}^\mu
   &=&
   \lambda_{21}^{-1}
   \Big( \left| \JCCvec{k}_{2,c} \right|, ~ \JCCvec{k}_{2,c} \Big) ,
\end{eqnarray}
with
\begin{equation}
    \lambda_{21} =
        \frac{ \left| \JCCvec{k}_{2a} + \JCCvec{k}_{2b} \right| 
               + \left| \JCCvec{k}_{2,c} \right|
             }
             {Q} ~.
\end{equation}
It is readily verified that $k_{a'}^\mu=k_a^\mu$ and $k_{b'}^\mu=k_b^\mu$.
This is the transitivity property: one gets the same momenta by any
appropriate sequence of projections.  The proof in general is made by
observing that each projection amounts to scaling all the relevant spatial
momenta by a factor.  The energies are defined by requiring the particles
to be massless, and the scaling factor is defined by requiring the total
energy always to be $Q$.

\paragraph*{Region $R_3$}
Just as with region $R_2$, I define
\begin{equation}
\label{eq:Gamma.3}
   \Gamma_3 = V_3 ~ T_{R_3} \!\left( \Gamma - \Gamma_1 \right) ,
\end{equation}
and I add it to the list of previously generated terms to obtain
$\Gamma_1 + \Gamma_2 + \Gamma_3$.  

But now a new problem appears when we try to show that $\Gamma_1 + \Gamma_2 +
\Gamma_3$ is equal to $\Gamma$ up to a non-leading term near {\em both} the
singular surfaces $R_2$ and $R_3$.  We have already seen that the sum
of two of the terms, $\Gamma_1+\Gamma_2$, is close to $\Gamma$ near $R_2$ --- see
Eq.\ (\ref{eq:R2.decomp}).  Of course, it is also true that the sum of
the other pair of terms, $\Gamma_1+\Gamma_3$, is close to $\Gamma$ near $R_3$.
But what we need is a result for the sum of all three terms.  This
extra complication is associated with the ``overlapping divergence''
structure in the relations between the regions, as symbolized in
Figs.\ \ref{fig:relation} and \ref{fig:region.geometry}.

What we need to show is that the sum of the three terms is a good
approximation in a neighborhood of both $R_2$ and $R_3$ (and of $R_1$),
i.e.,
\begin{equation}
\label{eq:R3.decomp}
   \Gamma = ( \Gamma_1 + \Gamma_2 + \Gamma_3)
      \left[ 1 + O(M^2/Q^2) \right] ,
\end{equation}
where now $M$ is the minimum of: (a) the mass of the pair of particles
1 and 2, and (b) the mass of the pair of particles 3 and 4.

The arguments I have used are sufficient provided that I prove the
additional results
\begin{itemize}
\item $\Gamma_2$ is power suppressed in region $R_3$:
    it is $O(\Gamma M_{12}^2/Q^2)$.
\item $\Gamma_3$ is power suppressed in region $R_2$: 
    it is $O(\Gamma M_{34}^2/Q^2)$.
\end{itemize}
These can be summarized by saying that each term is power-suppressed when
we approach a region that overlaps the design region of the given term.

Consider first $\Gamma_2$, whose definition is given in Fig.\ 
\ref{fig:Gamma.2}, or equivalently Eq.\ (\ref{eq:Gamma.2}).  It is
designed with region $R_2$ in mind, which is the region where the
lower two external lines ($p_3$ and $p_4$) are collinear.  That is, it
is concerned with a neighborhood of the line labeled $R_2$ in Fig.\ 
\ref{fig:region.geometry}.  Now $\Gamma_2$ contains a subtraction which
cancels the leading contribution from the smaller region $R_1$.  That
is, there is a power suppression, $O(M_{12}^2/Q^2)$ relative to the
size of $\Gamma$, when the upper two external lines become collinear.  This
suppression appears as one moves along the horizontal line in Fig.\ 
\ref{fig:region.geometry}.  

Now observe that this suppression holds over the whole region of phase
space, rather than just only around $R_2$, the horizontal axis in Fig.\ 
\ref{fig:region.geometry}.  The reason is simply that in the definition
Fig.\ \ref{fig:Gamma.2} of $\Gamma_2$, the dependence on $M_{34}$ is completely
factored out.  The cancellation that gives the $O(M_{12}^2/Q^2)$ estimate
works independently of the value of $M_{34}$, i.e., independently of
whether the $(34)$ pair is collinear.  To put it another way, the
application of the $T_{R_2}$ operation has put the $k_{2c}$ line in Fig.\ 
\ref{fig:region.geometry} exactly massless and on-shell, so it is as if
the $(34)$ pair were exactly collinear.

We therefore find that 
\begin{equation}
\label{eq:special.suppression}
    \Gamma_2 / \Gamma = O(M_{12}^2/Q^2) .
\end{equation}
Observe that when we integrate $\Gamma_2$ over all final states, the
power law suppression is worsened by a logarithm that arises from the
integral over the mass of the (3,4) pair.

The construction of $\Gamma_3$ is precisely such that 
$\Gamma_1 + \Gamma_3$ gives a good approximation to $\Gamma$ when $(12)$
are collinear, i.e., when $M_{12}$ is small.  Hence 
$\Gamma = (\Gamma_1 + \Gamma_2 + \Gamma_3) [1 + O(M_{12}^2/Q^2)]$.

Exactly similarly, $\Gamma_3$ is proved to be suppressed in region
$R_2$, and so
$\Gamma = (\Gamma_1 + \Gamma_2 + \Gamma_3) [1 + O(M_{34}^2/Q^2)]$.
We have two error estimates that are both valid, so the strongest
provides the best estimate, which is Eq.\ (\ref{eq:R3.decomp}).

These new results, as well as the suppression of $\Gamma_2$ and
$\Gamma_3$ in region $R_1$, are specific examples of a general result, to be
proved later in full generality:
\begin{quote}
   For any graph $\Gamma$ and any of its regions $R$, $C_R(\Gamma)$ is power
   suppressed in neighborhoods of all regions $R'$ except those 
   for which $R' \geq R$.
\end{quote}
Thus $C_R(\Gamma)$ gives a leading contribution in $R$ itself, as it should,
and in bigger regions.  Its very construction ensures that it is
suppressed in smaller regions.  We have now seen in an example that it is
also suppressed in other regions $R'$ that merely intersect with $R$ (or
do not intersect it at all).

\paragraph*{Region $R_4$}
The contribution associated with the purely ultra-violet region can
now be constructed as
\begin{eqnarray}
   \Gamma_4 &=&
       V_4 ~ T_{R_4} \!\left(
          \Gamma - \Gamma_1 - \Gamma_2 - \Gamma_3
       \right) .
\end{eqnarray}
The veto factor must cut out both of the extreme collinear regions, and we
can define it as
\begin{equation}
  \label{eq:V4}
  V_4 = V(\widetilde M_{12} / m ) ~ V(\widetilde M_{34} / m ) ,
\end{equation}
where $V$ is the elementary veto function defined in Eq.\ 
(\ref{eq:veto.function}).

\paragraph*{Sum}
With the use of all the results I have derived, it is easy to show
that the sum of all the terms $\Gamma_1 + \Gamma_2 + \Gamma_3 + \Gamma_4$ gives a good
approximation to $\Gamma$ in all the regions, so that we obtain Eq.\ 
(\ref{eq:Gamma.decomp}).  Observe that all these cancellations work
point by point in the momenta of the final-state particles.  Thus they
happen independently of the weighting function $W$, and thus
independently of which cross section we wish to compute.  This is in
contrast to the ``analytic'' methods of calculation in perturbative
QCD, which treat the subtractions as delta functions.  In the analytic
methods, the power suppression of the error only occurs after an
integral over final states with a collinear-safe weight function.


%% file: phi3-proof.tex

\section{Proof of factorization}
\label{sec:factorization.proof}

In this section, I will give the actual proof of the factorization
theorem.  The theorem is of a form that allows the construction of a
corresponding Monte-Carlo algorithm in Sec.\ 
\ref{sec:monte.carlo.algorithm}.  As is usual in such proofs, the details can
be tedious to work through unless one has a clear overall view of the
ideas and aims.  These are all based on experience with the examples
presented earlier.

So first, in Sec.\ \ref{sec:factorization.overall}, I will give an overall
view: One starts from the characterization of the possible regions that
are important, and then defines corresponding contributions with
subtractions to prevent double counting.  These individual contributions
are of such a form that summing over all possibilities gives a suitable
factorization formula.

The subsequent technical work is to show mathematically that the
subtraction terms do indeed perform the job that they are constructed to
perform.  This is done by constructing suitable estimates of graphs and
subgraphs before and after subtraction.

\subsection{Overall view}
\label{sec:factorization.overall}

As we have already seen, the leading regions for graphs are characterized
by a decomposition into subgraphs: a hard subgraph where all the internal
momenta are highly virtual, and two or more jet subgraphs.  So for a
leading region $R$ of a graph $\Gamma$, we can symbolize the decomposition
as
\begin{equation}
  H_R \times J_{R1} \cdots \times J_{RN_R} ,
\end{equation}
where $N_R$ is the number of jet subgraphs for the region.  Analytically,
the region is a neighborhood of a particular pinch-singular surface (PSS),
as shown by Libby and Sterman \cite{pc}, and the decomposition into
subgraphs corresponds to a factorization of the integrand.  

For each leading region we construct an approximation to the integrand by
applying a suitable Taylor expansion to the hard subgraph:
\begin{equation}
\label{eq:TR.fact}
  T_R\Gamma = T_R(H_R) \times J_{R1} \cdots \times J_{RN_R} ,
\end{equation}
Just as in the examples, the Taylor expansion is used to approximate the
hard subgraph by a massless on-shell partonic quantity and hence to
correctly approximate the momentum integrals by a suitable factorizable
form.  This is what enables convenient calculations to be done: after the
summation over graphs, the hard subgraph can be calculated independently
of the hadronization of the produced partons.

As we saw, we must avoid double counting, so the contribution for leading
region $R$ is obtained by applying the above manipulations not to the
original graph but to the original graph minus the previously constructed
terms for smaller leading regions.  In addition, we found it convenient to
insert a veto factor that excludes small neighborhoods of the massless
collinear configurations of the Taylor-expanded hard subgraph.  So we
define
\begin{equation}
\label{eq:CR.def}
   C_R(\Gamma) = V_R ~ T_{R} 
              \!\left( \Gamma - \sum_{R'<R} C_{R'}\Gamma \right) .
\end{equation}
This equation gives a recursive definition of the contribution $C_R(\Gamma)$
for region $R$, since it involves only the same quantity for smaller
regions.  The recursion starts with the minimal region(s) $R_{\rm min}$,
i.e., those that contain no smaller regions.  For these
\begin{equation}
   C_{R_{\rm min}}\Gamma = T_{R_{\rm min}} \Gamma .
\end{equation}
Hence Eq.\ (\ref{eq:CR.def}) is a valid definition of $C_R$.

The veto factor can be defined as a product of an elementary veto
functions:
\begin{equation}
  \label{eq:VR}
  V_R = \prod_{j} V(\widetilde M_j / m) .
\end{equation}
Here the product is over all combinations of external lines of the hard
subgraph of $R$, and $\widetilde M_j$ is the invariant mass of these lines
in the massless approximation.  The elementary veto function $V$ is
defined in Eq.\ (\ref{eq:veto.function}).  Although there is an
arbitrariness in the choice of the veto factors, the precise choice will
not affect the physics, as we have already observed,

Now follows a very critical point.  The double-counting subtractions are
applied on smaller leading regions than $R$ and thus to leading regions
whose hard subgraph is strictly smaller than $H_R$.  Moreover the Taylor
expansions are always applied to the hard subgraphs, with the jet factors
being left completely unaltered, as is appropriate for quantities that are
ultimately to be regarded as unknown non-perturbative quantities.  Hence
the subtractions are applied inside the hard subgraph for $R$, and we can
write
\begin{equation}
\label{eq:CR.from.H.J}
  C_R(\Gamma) =
  V_R ~ T_R \! \left( H_R - \sum_{R'<R} C_{R'}H_R \right)
  \times  J_{R1} \cdots \times J_{RN_R} .
\end{equation}

The argument now falls into the general pattern already explained in
Sec.\ \ref{sec:proof.summary}, and we simply need to explain what are
the jet factors $J_i$ and the hard factor $\hat{H}_N$ in Eq.\ 
(\ref{eq:basic.factorization}).  Their definitions follow from the
definition of $C_R(\Gamma)$.  The jet factors $J_i$ are (sums over)
unapproximated jet subgraphs, while the hard factor is obtained by
summing the hard-scattering factor in Eq.\ (\ref{eq:CR.from.H.J}) over
all the relevant graphs:
\begin{equation}
  \label{eq:Hhat}
  \hat{H}_N = \sum_{\rm graphs}   
             V ~ T \! \left( H - \sum_{R'<\infty} C_{R'}H \right) .
\end{equation}
The sum is over all (cut) graphs $H$ for the cross section which have
$N$ outgoing parton lines.  The unsubscripted $T$ denotes the Taylor
operator applied to the whole graph, i.e., that the whole graph is
treated as a hard graph, and similarly for the veto factor $V$.  The
sum over regions $R'$ is over all non-singular regions of $H$.  Thus
$\hat{H}_N$ is the properly Taylor-expanded and subtracted hard
subgraph with $N$ external parton lines.

So far, we have merely made definitions that generalize the structures
seen in examples.  The factorized form on the right-hand-side of Eq.\ 
(\ref{eq:basic.factorization}) is a trivial consequence of the
structure of the Taylor expansion operator that we have already
defined.  The non-trivial part of the proof of the factorization
theorem is to prove that the remainder term is indeed power
suppressed.  What we need to show is that, to leading power, each
individual graph is indeed the sum over the defined contributions for
each of its leading regions:
\begin{equation}
  \Gamma = \sum_R C_R(\Gamma) + \mbox{non-leading power} .
\end{equation}
That is, we must derive an appropriate estimate for the non-leading-power
term.  To employ the factorization theorem in calculations, certain
relatively simple subsidiary results will also be needed:
\begin{enumerate}
\item Infra-red and collinear safety of the subtracted hard scattering
  factors $\hat{H}$.
\item Infra-red safety of the {\em integrated} jet factors.
\item Renormalization-group-like equations for the scale dependence of the
  hard-scattering factors and the jets factors.
\end{enumerate}
See Thm.\ \ref{thm:IR.safety.H} and Secs.\ \ref{sec:IR.safety.int.jet}
and \ref{sec:RG}.

The actual proofs will be relatively short.  What will take the bulk of
the space will be a series of definitions that provide a precise language
for describing the various leading regions, their relations and the
accuracy of the approximations.  The main elements are:
\begin{enumerate}
\item The structure of the subtractions for each region.  This is an
  obvious generalization from the examples, and the structure is common to
  many related problems, for example the construction of renormalization
  counterterms.
\item A precise characterization of the regions to be discussed.
\item A quantification of the errors.  This needs a method of quantifying
  how close a point is to the defining surfaces of the various regions
  involved, which leads us to make a definition of the concept of a
  distance to a singular surface.
\item The actual construction of the error estimates.  This involves
  considerations of the relative sizes of the virtualities of different
  lines, which will be characterized by the distances to singular
  surfaces. 
\end{enumerate}
One tricky point is that we will want to prove a suitable error estimate
for a given graph given that the subtractions inside hard subgraphs
actually work.  The subtractions work if corresponding error estimates are
available for the hard subgraphs.  The simplest and most obvious proofs
are made with the external lines of a graph on-shell.  But when these same
graphs occur as subgraphs of bigger graphs, we will in fact need the
estimates for graphs with off-shell external lines, and the notion of
distance to a singular surface needs to be carefully defined to allow for
this possibility.

There will also be two views of the leading momentum regions.  One will be
a geometrical and topological view, of the kind illustrated by Fig.\
\ref{fig:region.geometry}, which is symbolic of a momentum space of quite
large dimension.  In this view the key ideas are the distances to singular
surfaces and the relations between the singular surfaces defined by set
theoretic inclusion and intersection.  The second view relates these
regions to Feynman graphs, the sizes of the hard and jet subgraphs for
different regions, as symbolized in Fig.\ \ref{fig:Njet}.  The error
estimates are given by relating virtualities of lines in Feynman graphs to
the distances to singular surfaces.

\subsection{General formula for single graph}
\label{sec:single.graph}

The structures seen in the examples suggest the general result.  To
make the argument self-contained and precise, I will give the results
and proofs as a sequence of definitions and theorems.  The first few
recapitulate results from the previous sections.

\begin{theorem}
  The non-ultra-violet contributions to the cross section $\sigma[W]$ arise
  from neighborhoods of the pinch singular surfaces (PSS) for the
  massless theory.  Moreover the leading-power contributions in our
  model correspond to the regions symbolized in Fig.\ \ref{fig:Njet}.
  The corresponding PSS have groups of jet lines consisting of exactly
  parallel massless momenta.
\end{theorem}
\begin{proof}
This was proved by Libby and Sterman \cite{pc} in their fundamental
work on the analysis of the leading regions.  
A more recent account of the power-counting is given in \cite{newer.pc}.

This result applies to both real and virtual corrections to the process.
In the case of loop diagrams, it is only necessary to consider regions
where the (multi-dimensional) contour of integration is trapped near
on-shell configurations for the virtual lines; otherwise the contour can
be deformed.  This accounts for the requirement that the singular surfaces
correspond to a pinch.
The momenta for external on-shell lines are, of course, necessarily
pinched. 
\end{proof}

\begin{definition}
  The regions $R$ for a graph $\Gamma$ are specified by the corresponding
  PSS, with the inclusion of the limit points of the surfaces.  In
  addition I define one other region, the purely ultra-violet region,
  whose associated surface is the surface where all the external lines of
  the graph are massless.  I define a singular region to be one that
  corresponds to a non-trivial PSS; the only non-singular region is the
  ultra-violet region.
\end{definition}

This definition was explained and motivated by means of an example in
Sec.\ \ref{sec:harder.example}.  Recall that it was convenient to code
each region by a set of points that is not the region itself but a
related surface of singularity in a massless theory.  The above
definition is simply the natural generalization of the example.

One odd-looking feature is the definition of the surface associated with
the purely ultra-violet region.  For the example of the 4-parton graph of
Fig.\ \ref{fig:4jet}, this was $R_4$, and in Sec.\ \ref{sec:examples} its
associated surface was defined to be the whole of the space of final-state
momenta.  But now we will also want to be able to discuss this graph when
it is the hard subgraph for a particular region of a bigger graph.  We
will be using a Taylor expansion operator that replaces the external
momenta of the hard subgraph by massless momenta.  Therefore it is more
appropriate to consider the space of momenta to be the space of all
momenta (on-shell or not), and then the surface associated with $R_4$ is
most naturally the surface where all the external momenta of the graph are
massless.  This surface then has an obvious embedding into a bigger space
of momenta when the graph is a subgraph of a bigger graph.  The above
definition simply generalizes this idea to all graphs.

The inclusion of the limit points is an important technicality that lets
us define the ordering of regions simply as set-theoretic inclusion.
Physically the inclusion of limit points corresponds to including smaller
regions in the definition of the defining surfaces.  Initially, I found
this to be contrary to my intuition.  For example, in the case of Fig.\ 
\ref{fig:4jet} one might suppose the region $R_2$ where $p_3$ and $p_4$
are parallel should exclude the region where in addition $p_1$ and $p_2$
become parallel.  The approximation to this region provided by applying
$T_{R_2}$ certainly loses accuracy in the smaller region since it is no
longer valid to neglect masses in the hard subgraph.  However, the
mathematics runs much more smoothly if one defines the surface that codes
a particular region to include the surfaces that code each of the smaller
regions.

\begin{definition}
  An ordering of the regions is defined by set-theoretic inclusion: $R_1 <
  R_2$ means that the defining surface for $R_1$ is strictly smaller than
  the defining surface for $R_2$.
\end{definition}

It is convenient to make explicit the concept of a leading region:
\begin{definition}
  A leading region is a region that gives a leading-power contribution to
  the cross section.  A non-leading region is a neighborhood of a PSS that
  gives a non-leading-power contribution.
\end{definition}

\begin{theorem}
  {\em Feynman-graph interpretation of ordering for leading regions:} Let
  $R_1$ and $R_2$ be two leading regions for a graph $\Gamma$.  Then
  $R_1<R_2$ is equivalent to saying that the hard subgraph, $H_1$, for
  $R_1$ is smaller than the hard subgraph, $H_2$, for $R_2$.
  Correspondingly, when momenta approach the smaller region $R_1$, some
  internal lines of $H_2$ have propagators with small denominators
\end{theorem}
\begin{proof}
  This is an elementary consequence of the meaning of the collinear
  kinematics of the singular surfaces:  If $R_1$ is smaller than $R_2$,
  then more lines of the graph have low virtuality in region $R_1$ than in
  region $R_2$.  Thus fewer lines are far off-shell and the hard subgraph
  is smaller.  
\end{proof}

\begin{definition}
  The approximation associated with a region $R$ is notated by
  $T_R\Gamma$.  In the case of a leading region, the definition of $T_R\Gamma$
  was given in Sec.\ \ref{sec:kinematics} at Eq.\ 
  (\ref{eq:main.approx}).  Observe that $T_R\Gamma$ is defined to extract
  the first term in the Taylor expansion in the masses of the external
  lines of the hard subgraph for the region, and then to use massless
  phase space for these external lines.  (So it is not a naive Taylor
  expansion.) For a non-leading region, we define $T_R\Gamma = 0$.
\end{definition}

\begin{theorem}
  {\em Basic properties of $T_R\Gamma$:} The approximation $T_R\Gamma$ for a
  leading region $R$ has the form of a hard subgraph $H(\hat{l})$, with
  massless external lines, times a product of jet factors, as in Eq.\ 
  (\ref{eq:main.approx}) or (\ref{eq:TR.fact}).  The approximation can
  be expressed in terms of the original phase-space measure, as in
  Eq.\ (\ref{eq:phase.space.transformation}), with the aid of a
  Jacobian factor.  The Jacobian, Eq.\ (\ref{eq:Jacobian.6.dim}), is
  analytic in a neighborhood of the surface $R$.
\end{theorem}
\begin{proof} 
  These are immediate consequences of the definition, and all the
  necessary discussion is in Secs.\ \ref{sec:approximation} to
  \ref{sec:jacobian}.
\end{proof}

\begin{definition}
  Let $\Gamma$ be a general graph for the cross section.  Then the
  contribution $C_R(\Gamma)$ associated with a region $R$ is defined
  recursively by Eq.\ (\ref{eq:CR.def}).  This is the appropriate
  approximation applied to the original graph minus the contributions
  associated with smaller leading regions.  (Observe that for a
  non-leading region $C_R(\Gamma)=0$, so that only leading regions need be
  considered explicitly.)
\end{definition}

It is convenient to define a related quantity, which is the original
graph minus the contributions associated with smaller regions:
\begin{definition}~
\begin{equation}
\label{eq:CRbar.def}
   \overline{C}_R(\Gamma) = \Gamma - \sum_{R'<R} C_{R'}(\Gamma) .
\end{equation}
\end{definition}
A trivial consequence is that 
\begin{equation}
\label{eq:CR.from.CRbar}
   C_R(\Gamma) = V_R ~ T_R \overline{C}_R(\Gamma) .
\end{equation}

A related definition is:
\begin{definition}
\label{def:Gammabar}
  For a graph $\Gamma$, we let $\overline\Gamma$ be the graph with subtractions
  made for all of its leading singular regions:
  \begin{equation}
    \label{eq:Gamma.bar}
    \overline\Gamma = \Gamma - \sum_{R<\infty} C_R \Gamma .
  \end{equation}
  (Only leading regions contribute, since otherwise $C_R \Gamma = 0$.)
  The notation $R<\infty$ means that the purely ultra-violet region is
  {\em  not} included in the sum, i.e., that only singular regions are
  included in the sum.  
\end{definition}
For example, the singular regions for the graph of Fig.\ \ref{fig:4jet}
are $R_1$, $R_2$ and $R_3$.  In both of these two definitions, the bar
over a quantity means that all subtractions are made except for the
outermost one.  

The notation defined by Def.\ \ref{def:Gammabar} gives simple
expressions for quantities we interested in.  For example, applied to
the hard subgraph for a particular region, it lets us write the
contribution associated with the region as
\begin{equation}
\label{eq:CRbar.Hbar}
   C_R\Gamma = \prod_i J_{Ri} ~ V_R ~ T \overline{H}_R ,
\end{equation}
where $H_R$ is the hard subgraph for a leading region $R$, $J_{Ri}$
are its jet subgraphs, and $\overline{H}_R$ is obtained from $H_R$ by using
Eq.\ (\ref{eq:Gamma.bar}).

It is perhaps worth reminding ourselves that $T \overline{H}_R$ is obtained
from $\overline{H}_R$ by
\begin{enumerate}
\item Setting the external lines massless and on-shell, and setting all
  the internal masses to zero, and
\item Inserting a Jacobian factor corresponding to the transformation
  between the exact momenta and the approximated massless on-shell
  momenta. 
\end{enumerate}

I will show that the sum of the contributions for the different leading
regions gives the leading power behavior of $\Gamma$, so it is useful to
make the following definitions:
\begin{definition}
\label{def:rem.asy}
The asymptotic part of $\Gamma$ is
\begin{equation}
\label{eq:def.Asy}
   \mathop{\rm Asy}\Gamma = \sum_R C_R(\Gamma).
\end{equation}
Here the sum over leading regions includes not only the regions
corresponding to leading singularities of the massless graphs, but
also the purely ultra-violet region, where all the external lines of
$\Gamma$ are at wide angles.

The remainder is defined by any of the equivalent formulae
\begin{equation}
\label{eq:def.Rem}
   \mathop{\rm Rem}\Gamma
    = \Gamma - \mathop{\rm Asy}\Gamma
    = \Gamma - \sum_R C_R(\Gamma)
    = (1- VT)\overline{\Gamma} ,
\end{equation}
where the unsubscripted $T$ in the last form denotes the Taylor
operator applied to the whole graph, i.e., where the whole graph is
treated as a hard graph, and similarly for the veto factor $V$.
\end{definition}
The main work of the rest of this section is to prove that the asymptotic
part of $\Gamma$ is indeed a good approximation to $\Gamma$, i.e., that
$\Gamma = \mathop{\rm Asy}\Gamma \times \left[ 1 + O(m^2/Q^2) \right]$
plus power-suppressed terms from non-leading regions, or, equivalently,
that the remainder is power-suppressed: 
$\mathop{\rm Rem}\Gamma = \Gamma \times O(m^2/Q^2)$, again up to
power-suppressed contributions from non-leading regions.

The reason for making an explicit definition of $\mathop{\rm Rem}\Gamma$
is that it unifies the treatments of leading and non-leading regions:
We will prove that in all regions $\mathop{\rm Rem}\Gamma$ is power
suppressed.

In the following theorems our aim will be to give estimates of the
sizes of various quantities and of the errors in approximations, just
as we did in specific examples.  Estimates will be propagated through
further manipulations.  The actual integrands of Feynman graphs have
an enormous range of sizes; for example near a collinear configuration
some propagator denominators get small.  This happens in such a way
that an integral over the region gives merely a logarithmic
enhancement in the cross section.  To avoid having to explicitly state
the sizes involved every time, it is convenient to use a notation
$\|\gamma\|$ for a previously derived estimate of some graph or subtracted
graph $\gamma$.  An example of such an estimate is Eq.\ 
(\ref{eq:A.minus.A1}), where for further manipulations in a general
context, it would be useful to write $\|A-A_1\| = \|A\| M_{12}^2/Q^2$.
The estimate of $A$, the unsubtracted graph itself, could simply be
defined as the absolute value of $A$: $\|A\| = |A|$.  Note that these
estimates are functions of all the relevant kinematic variables and
that we will require the estimates to be positive definite.  In a
theory with spin, there could be ``accidental'' zeros in the numerator
factors; such accidental zeros should be removed from the definition
of $\|\gamma\|$.

In addition it is convenient to define $L(\Gamma)$ to be a value that is
leading twist in all singular regions of $\Gamma$---it differs from
$\|\Gamma\|$ by being multiplied by a suitable dimensionless enhancement
factor for each non-leading singular region.

Error estimates such as these involve some notion of distance to a
singular surface, and we must of course define what we mean by a
distance.  Since there is no metric on the space of momenta over which
we integrate, the definition is not unique, but for the purposes of
error estimation, the precise choice of definition will not matter.  A
suitable definition can be constructed by starting with a choice of
coordinates:
\begin{definition}
  For each singular surface $R$, we choose ``parallel'' and ``normal''
  variables.  This means that we change the integration variables to have
  the form $(k_\parallel,k_\perp)$, where $k_\parallel$ is used to
  parameterize the surface $R$, and $k_\perp$ are normal variables.  Thus
  the surface $R$ is $k_\perp=0$, while the region corresponding to $R$ is
  just a neighborhood of $k_\perp=0$.
\end{definition}
There is considerable freedom in the choice of these variables.  One
natural set for $k_\parallel$ consists of:
\begin{itemize}
\item The longitudinal momenta of the lines in the jet subgraphs,
  divided by $Q$. 
\item The total energy of each jet, divided by $Q$.
\item The angles between the external lines of the hard subgraph.
\end{itemize}
These variables are arranged to be dimensionless and to have a range
of values of order unity; this will aid the use of dimensional
counting to compute the dependence on $Q$.  

The remaining variables, $k_\perp$, parameterize the deviation of momenta
from the exactly singular configuration for the region.  Given that we
will wish to vary the virtualities of the external lines of a graph,
an appropriate definition for the internal variables of a jet subgraph
is given by the following steps:
\begin{itemize}
\item As in Sec.\ \ref{sec:kinematics}, let the total jet momentum for the
  subgraph be $l_j^\mu$ and let the final-state momenta be $p_{j,i}^\mu$. 
\item Use light-front coordinates for the momenta in the jet, in the
  overall center-of-mass frame, with the $z$ axis being chosen as the
  direction of the jet's total spatial momentum $\JCCvec{l}_j$.  The $+$
  and $-$ momenta of the final-state lines are $p_{j,i}^{\pm} =
  (p_{j,i}^0\pm p_{j,i}^z)/\sqrt{2}$, and there are also four components
  of transverse momentum for each $p_{j,i}^\mu$.  The axes are
  different for each jet subgraph.
\item The fractional $+$ momenta, e.g., $p_{j,i}^+/l_j^+$, can be chosen
  to be among the parallel variables for the singular surface $R$, as
  listed earlier, in place of the longitudinal momenta of the lines in the
  jet subgraph.
\item The $-$ momenta and the transverse momenta form the normal
  variables.   (This applies both to the final-state momenta of the jet
  and to any loops.)
\end{itemize}
These variables have convenient properties under boosts along the jet
direction. 

We will also need to consider non-leading regions.  These differ from
leading regions by the possibility of extra lines beyond the minimum
of two joining a jet subgraph to the hard subgraph and by the
possibility of a soft subgraph connecting to the jet and/or hard
subgraphs --- see \cite{Analytic.Review,pc}.  The above definitions
apply equally to the hard and jet subgraphs of non-leading regions,
and we need only supplement them by adding the soft momenta to the
list of normal variables $k_\perp$.

To define the notion of distance, we make the following definition:
\begin{definition}
  We write the normal variables $k_\perp$ in terms of a single radial
  variable $\lambda$ and some ``angular'' variables $\tilde{k}_\perp$.  This
  can be done by writing $k_T = \lambda \tilde{k}_T$ and $k^- = \lambda^2
  \tilde{k}^-/Q$ for each internal jet momentum, and
  $k^\mu=\lambda\tilde{k}^\mu$ for each soft momentum.  Then we impose a
  suitable normalization condition on the scaled variables, for
  example, $\sum_{\mathrm{jet~}i}( \tilde{k}_{i,T}^2 +
  |\tilde{k}_i^-k_i^+/Q|) + \sum_{\mathrm{soft~}i,\mu}|\tilde{k}_{{\rm
      soft~}i}^\mu|^2= 1$.  Thus the normal variables $k_\perp$ have been
  expressed in terms of $\lambda$, and all but one of the scaled momentum
  components\footnote{ Observe the difference between a subscript $T$
    and a subscript $\perp$.  In this paper, the notation $T$ will be
    used for the transverse components of ordinary Lorentz vectors, as
    usual, whereas the notation $\perp$ will only be used to indicate the
    very different concept of a normal variable associated with a
    region of the space of all the momenta of a graph.  
  }.  The scaling, for reasons explained below, is different between
  the transverse and $-$ components of jet momenta, and  $\lambda$ is
  arranged to have the dimensions of mass.
\end{definition}
Since there is no natural metric on a space of several momentum variables,
this choice of variable is not unique.  But the details do not matter.  It
is the scaling properties with respect to $\lambda$ and $Q$ that will
dominate our discussion.  The normalization condition on the
$\tilde{k}_\perp$ variables is not Lorentz invariant; this is satisfactory
since the power-counting method and the derivation of the singular regions
rely on the use of a preferred rest frame.

The different scaling of the $-$ and transverse components of jet
momenta is chosen so that the square of a jet momentum
$k^2=2k^+k^--k_T^2$ simply scales homogeneously, proportional to
$\lambda^2$.  This will be convenient in constructing error estimates that
rely on the virtualities of the lines of a graph.  Observe further
that the scaled variables are dimensionless with a range of order
unity, and that $\lambda$ ranges up to $Q$.

For an example of the meaning of the variable $\lambda$, consider the
graph of Fig.\ \ref{fig:4jet}.  There are 14 variables of integration, 
which we split into 12 angular variables and the masses, $M_{12}$
and $M_{34}$, of two pairs of particles.  The graph is independent of the
12 angles, so we only need to consider the two pair masses.

In region $R_1$ we can define $\lambda_1 = \sqrt{M_{12}^2 + M_{34}^2}$.
Since we have agreed to ignore the angles, there are no parallel
variables for this region, and the normal variables consist of
$M_{12}$ and $M_{34}$.

In region $R_2$, we can define $\lambda_2 = M_{34}$, and this is the only
normal variable.  There is one parallel variable, $M_{12}/Q$.  

For the general case we can write the integral over a graph $\gamma$ as
\begin{equation}
\label{eq:parallel.perp}
  \int dk \, \gamma
=
  Q^2 \int dk_\parallel \int d\tilde{k}_\perp 
  \int \frac{d\lambda}{\lambda}  \, 
  \tilde{\gamma}( \lambda/m, \lambda/Q, k_\parallel, \tilde{k}_\perp ) .
\end{equation}
Here we have taken advantage of the fact that the variables $k_\parallel$ and
$k_\perp$ are dimensionless.  The overall factor $Q^2$ corresponds to the
dimension of the integrated graph.  The remaining part of the Jacobian
for the transformation of variables has been absorbed into the
integrand to give $\tilde{\gamma}$, which is dimensionless.  The
mass-shell delta functions for the final-state particles are treated
as part of the integrand.  The integrals over $k$, $k_\parallel$ and $\tilde
k_\perp$ are all multidimensional, of course.  

The power-counting results of \cite{pc} tell us how $\tilde{\gamma}$
behaves as $Q\to\infty$ with $\lambda$, $k_\parallel$, and
$\tilde{k}_\perp$ fixed.  The leading power occurs for the regions
specified by Fig.\ \ref{fig:Njet}, for which $\tilde{\gamma}$ goes to a
generally non-zero constant:
\begin{equation}
  \label{eq:leading}
  \mbox{Leading region:~~}
  \tilde{\gamma}( \lambda/m, \lambda/Q, k_\parallel, \tilde{k}_\perp ) 
  \to 
  \tilde{\gamma}( \lambda/m, 0, k_\parallel, \tilde{k}_\perp ) 
  \neq 0.
\end{equation}
All other regions give a negative power of $Q$:
\begin{equation}
  \label{eq:nonleading}
  \mbox{Non-leading region:~~}
  \tilde{\gamma}( \lambda/m, \lambda/Q, k_\parallel, \tilde{k}_\perp ) 
  \sim
  \left( \frac{ \lambda }{ Q } \right)^p ,
\end{equation}
where the power $p$ is, for the case of our model, at least 2.  

Now $\lambda$ ranges from 0 to order $Q$.  When mass-shell constraints are
allowed for, the minimum value is restricted to be of order $m$.  When we
perform the integral over all $\lambda$ in a leading region, a logarithm
results from the region $m \ll \lambda \ll Q$, whose coefficient is
obtained by neglecting the mass in $\tilde{\gamma}$:
\begin{equation}
  \mathrm{Logarithm} =
  \int_m^Q \frac{ d\lambda }{ \lambda }
  \tilde{\gamma}( \infty, 0, k_\parallel, \tilde{k}_\perp ) .
\end{equation}
Further logarithms result from the integrals over $k_\parallel$ and
$\tilde{k}_\perp$.

For a non-leading region, the integral over $\lambda$ is roughly
\begin{equation}
    \int_m^Q \frac{ d\lambda }{ \lambda } 
   \left( \frac{ \lambda }{ Q } \right)^p ,
\end{equation}
which is dominated by $\lambda \sim Q$: the contribution from $\lambda
\sim m$ is of order $(m/Q)^p$.  

The primary operation in computing the asymptotics of a graph is the
application of the Taylor expansion operator $T_R$ for each region
$R$.  The definitions we have just made characterize this operator
essentially as the extraction of the leading term in an expansion in
inverse powers of $Q$, for large $Q$.  The errors in the use of this
expansion are associated with the neglect of $\lambda$ and $m$ with respect
to virtualities in the hard subgraph for the region $R$.  For a
generic value of $k_\parallel$, the virtualities in $H_R$ are of order
$Q^2$, and the errors have relative size $\lambda^2/Q^2$.  But when $k_\parallel$
approaches values associated with smaller singular surfaces, some of
the virtualities get smaller, and the errors get larger.  To quantify
this, it is convenient to define the distance of a point of momentum
space from the defining surface of a region:
\begin{definition}
  Given a choice of normal and parallel variables for each $R$ and of the
  overall scaling of the normal variables, we define the distance of a
  point $k$ of momentum space to the region $R$ as the value of
  $\lambda=\lambda(k,R)$.
\end{definition}

To estimate the errors in our expansion, etc., we need to relate the
distance to a singular surface to the virtualities of lines of a graph.

\begin{theorem}
\label{thm:virt.dist}
   Let $k$ be a point of the momentum space for a graph $\Gamma$. Then the
   smallest virtuality, $V_{\rm min}(k)$, of the internal lines of $\Gamma$
   is at least of order the minimum of $\lambda^2(k,R)$ taken over all
   singular 
   regions $R$ of $\Gamma$.  This means that the ratio 
   $\min_{R<\infty}\lambda^2(k,R)/V_{\rm min}(k)$ is bounded: it is less
   than a constant that is independent of $k$.
\end{theorem}
\begin{proof}
  If we restricted our attention to {\em leading} singular regions, the
  proof would be quite simple, since in our theory, the leading regions
  correspond to groupings of external lines into jets.  Then the distances
  to regions can simply be related, in order of magnitude, to the
  invariant masses of groups of external lines.  
  
  But complications arise since we will actually need to handle momenta in
  non-leading regions, and for them the scalings vary for different
  momentum components, and the relation to invariant masses is not so
  clear. 
  
  First, consider any particular singular region $R$.  For generic values
  of $k$, the virtualities of the hard subgraph are of order $Q^2$ and the
  virtualities of the jet lines are of order $\lambda^2(k,R)$, if the
  region is a leading region, and hence without a soft subgraph.  For
  non-leading regions there may be soft lines so that some jet lines get a
  virtuality of order $Q\lambda$.  Hence for generic momenta at a distance
  $\lambda$ from the region, the virtualities of internal lines of the
  graph are at least $\lambda^2$.  
  
  However, by varying $k_\parallel$ and $k_\perp$ we can generally get
  lower virtualities, than the generic estimate, at a given value of
  $\lambda$. 
  
  Now fix $k$ and find the singular region $R_m$ that minimizes $\lambda$.
  It may be that $k$ is a generic value for this region, in which case
  $V_{\rm min}(k) \gtrsim \lambda^2(k,R_m)$ and the theorem is proved.  We
  will need to prove that no other possibility arises.
  
  So suppose the opposite, that $V_{\rm min}(k) \ll \lambda^2(k,R_m)$.
  Then some lines of the graph have much lower virtualities than is given
  by the generic estimate.  Let us call these ``anomalous lines''.  Then
  we can make a {\em small} adjustment of $k$ to put the anomalous lines
  exactly on-shell.  But this corresponds to another singular region,
  $R_n$, say.  Since some of the jet or soft lines of the original region
  $R_m$ are in the hard subgraph of the new region, the new region is
  either larger or non-trivially overlaps with the previous one.  
  
  Because the anomalous lines of $R_m$, i.e., those with especially low
  virtuality, consist of all the jet and soft lines of the new region, we
  will be able to show that the distance to the new region $R_n$ is less
  than the distance to the previous regions, $R_m$, contrary to the
  definition of $R_m$.  Indeed, if $k$ is counted as a generic momentum
  configuration for the new region $R_n$, then $V_{\rm min}(k) \gtrsim
  \lambda^2(k,R_n)$, which is just this result.  If $k$ is not generic,
  then we just repeat the above argument with $R_m$ replaced by $R_n$.  
\end{proof}

\begin{theorem}
\label{thm:accuracy}
{\em Accuracy of approximation given by $T_R$ for a leading region.}
Let $\gamma$ be a graph, and let $\|\gamma\|$ represent an estimate of its
size.  Suppose $k$ is a point which is a distance $\lambda$ from the
defining surface of one of its leading regions $R$, and suppose that
the minimum virtuality of the lines in the hard subgraph $H_R$ is
$V_{H, \rm min}$.  Then
\begin{equation}
\label{eq:TR.accuracy}
   \gamma - T_R \gamma = 
   \left\{
   \begin{array}{ll}

       O\!\left(
              \| T_R \gamma \|  
              \displaystyle \frac{ \max(\lambda^2,m^2) }{ V_{H, \rm min} }
       \right) 
       & {\rm if~} \max(\lambda^2,m^2) \lesssim V_{H, \rm min} ,
   \\[4mm]
       O\!\left(
              \| T_R \gamma \|  
       \right)
       & {\rm otherwise} .
   \end{array} 
   \right.
\end{equation}
The minimum virtuality $V_{H, \rm min}$ can be replaced by the square of
the distance to the nearest singular surface smaller than $R$, if such a
surface exists.  Let us call this distance $M_B(k,R)$.  If no smaller
singular surface exists, or if the hard graph is a lowest-order tree graph
so that it has no internal lines, then $M_B$ is to be replaced by $Q$, and
$V_{H, \rm min}$ by $Q^2$.
\end{theorem}
\begin{proof}
  The whole point of the definition of $T_R$ was to make a useful
  approximation to a graph valid as $\lambda\to0$, and this theorem
  formalizes how accurate this approximation is.
  
  Now $T_R\gamma$ is obtained by replacing the hard subgraph for the
  region $R$ by the first term in its expansion about the collinear
  limit for its external lines.  The jet subgraphs are completely
  unapproximated, so that they do not enter into the error estimate.
  The errors come from three sources: (a) the neglect of the external
  virtualities for the hard subgraph in comparison with the
  virtualities of its internal lines, (b) the neglect of internal
  masses in the hard subgraph, and (c) in the calculation of the
  massless phase space measure, the neglect of the external
  virtualities of the hard subgraph in comparison with $Q^2$.  
  
  The largest relative error comes from the internal lines of the hard
  subgraph that have the smallest virtualities; and these small
  internal virtualities arise precisely from neighborhoods of smaller
  regions than $R$.  The error arises from neglecting both $\lambda^2$ and
  $m^2$, so that we must insert the largest of these, i.e.,
  $\max(\lambda^2,m^2)$ in the numerator of Eq.\ (\ref{eq:TR.accuracy}).
  Since terms neglected in the propagator denominators have factors of
  external virtualities and mass-squared, and since the virtualities
  of the external lines are proportional to $\lambda^2$, the error term is
  quadratic, not linear, in $\lambda$ and $m$.
  
  When the minimum virtuality of the lines of the hard subgraph
  decreases below $\lambda^2$ or $m^2$, the approximation to $\gamma$ given by
  $T_R\gamma$ becomes totally inaccurate, and the error is simply
  estimated as $100\%$ of the bigger of $\gamma$ and $T_R\gamma$, which is
  normally the more singular quantity $T_R\gamma$.

  Thus we have proved Eq.\ (\ref{eq:TR.accuracy}).
  
  For the subsequent discussion, it is convenient to work in terms of
  the geometry of the momentum integration space and hence in terms of
  distances to singular surfaces.  Small virtuality lines in the hard
  subgraph are obtained by letting the parallel variables $k_\parallel$ for
  the region $R$ approach the defining surfaces for smaller singular
  surfaces. The surfaces are smaller, because the regions of momentum
  are more restricted: the external lines are already of low
  virtuality, and we are finding places where additional lines have
  low virtuality.  Then essentially the same argument that led to
  Thm.\ \ref{thm:virt.dist} shows that the minimum virtuality can be
  replaced by the square of the distance to the nearest smaller
  singular surface.  This quantity repeatedly appears later, so we
  give it a name $M_B^2(k,R)$.
  
  Of course, if there are no smaller singular surfaces, and, in
  particular, if the hard subgraph is a lowest-order tree graph, so that
  it has no internal lines, then the approximations involve the neglect of
  $\lambda^2$ and $m^2$ with respect to $Q^2$.  In this case we simply
  define $M_B^2$ to be $Q^2$.
\end{proof}

\begin{theorem}
\label{thm:accuracy.2}
{\em Accuracy of approximation given by $T_R$ times veto factor.}  
For any point $k$, 
\begin{equation}
\label{eq:TR.accuracy.2}
   \gamma - V_R T_R \gamma = 
   O\!\left(
          \| \gamma \|  \frac{ \max(\lambda^2,m^2) }
                      { \max(M_B^2, \lambda^2, m^2) }
    \right) .
\end{equation}
\end{theorem}
\begin{proof}
  This is the same theorem as Thm.\ \ref{thm:accuracy},
  except that $T_R\gamma$ has been multiplied by the veto function defined
  earlier, and that in the error estimate $T_R\gamma$ has been replaced by
  $\gamma$. 

  The veto factor multiplying $T_R\gamma$ ensures that whenever we
  get close to the massless collinear singularities of $T_R\gamma$, this
  quantity is replaced by zero.  In that case $\gamma - V_R T_R \gamma$ is just
  equal to $\gamma$, and a correct error estimate is just 100\% of
  $\gamma$, instead of the commonly larger $T_R\gamma$.  

  The complicated looking denominator in Eq.\ (\ref{eq:TR.accuracy.2}) 
  is just to combine the two cases in Eq.\ (\ref{eq:TR.accuracy}).
\end{proof}

\begin{definition}
  In Thm.\ \ref{thm:accuracy}, the quantity $M_B$ played an important
  role.  So it is useful to formalize it and to make some related
  definitions.  We consider a graph $\Gamma$ and a point $k$ of its
  momentum space.  Then
  \begin{itemize}
  \item $M_B(k, R)$ is the minimum distance to the defining surface of
    regions $R'$ that are smaller than $R$:
    $M_B(k,R)=\min_{R'<R}\lambda(k,R')$. If there is no smaller region,
    then $M_B$ is defined to be $Q$. 
  \item $M_B(k, \Gamma)$ (with the argument being the graph $\Gamma$ instead
    of a region $R$) is the minimum distance to a singular region of
    $\Gamma$. Thus, $M_B(k, \Gamma) \equiv \min_{R<\infty}\lambda(k,R)$.
  \item $M_{\rm ext}(k, \Gamma)$ is the maximum of the square roots of the
    virtualities of the external lines of $\Gamma$.  
  \end{itemize}
\end{definition}


The basic result that gives the error on our approximations is Thm.\ 
\ref{thm:accuracy}.  Its simplest application is to a graph $\gamma$ that
has no singular regions.  Then the only region is the ultra-violet
region, and the difference between $\gamma$ and $T_R\gamma$ only comes from
the neglect of external virtualities with respect to $Q^2$ in the
calculation of the phase-space Jacobian.  This is equivalent to the
observation that the value of $\lambda^2$ in Eq.\ (\ref{eq:TR.accuracy}) is
of order the maximum external virtuality, i.e., $M_{\rm ext}^2$.  If
the graph has on-shell external lines, then the error is of relative
order $m^2/Q^2$, and therefore $\mathop{\rm Rem}\Gamma=O(\|\Gamma\|m^2/Q^2)$.


To prove this result for graphs with non-trivial leading singular
regions, we will use an inductive (or recursive) approach, proving the
desired result on the assumption that suitable results for the errors
of smaller graphs have been derived.  The necessary results are
conveniently stated with the aid of the following definition:
\begin{definition}
\label{def:well.behaved}
  A graph $\Gamma$ is well-behaved if
  \begin{equation}
    \label{eq:well.behaved}
    \overline\Gamma 
    = O\!\left(
        L(\Gamma) \frac{ M_B^2(k,\Gamma) }{ Q^2 }
      \right)
  \end{equation}
  for all momenta $k$ whenever the external momenta of $\Gamma$, which
  (for this definition) are not necessarily on shell, obey 
  $M_{\rm ext}(k,\Gamma) < M_B(k,\Gamma)$.  Here, as defined by Eq.\ 
  (\ref{eq:Gamma.bar}), $\overline\Gamma$ is the graph after subtractions
  for leading singular regions have been made.  Also, as defined
  earlier, $L(\Gamma)$ is a quantity like $\|\Gamma\|$ modified so that it
  gives a leading contribution near every singular region.
  
  More informally, we are defining $\Gamma$ to be well-behaved if the
  subtracted graph $\overline\Gamma$ is power suppressed near every
  singular surface.

  An equivalent formulation can be made in terms of the distances to
  individual singular regions.  Thus $\Gamma$ is well-behaved if
  \begin{equation}
    \label{eq:well.behaved.R}
    \overline\Gamma 
    = O\!\left(
        L(\Gamma) \frac{ \lambda^2(k,R) }{ Q^2 }
      \right)
  \end{equation}
  for all regions $R$ and for all momenta $k$ whenever the external
  momenta obey $M_{\rm ext}(k,\Gamma) < M_B(k,\Gamma)$.
\end{definition}
The idea of this definition is that $\Gamma$ may have singular regions,
where it is substantially enhanced, and that without subtractions the
integrals near the leading singular regions have strong logarithmic
enhancements, which diverge if $m \to 0$.  The aim of the subtractions
that are used to define $\overline\Gamma$---see Eq.\ 
(\ref{eq:Gamma.bar})---is to provide a suppression in each of these
regions.  The above definition gives a criterion as to whether our
definitions of the subtractions have succeeded in their aim, and so we
will need to prove that all graphs are in fact well-behaved, in the
sense of this definition---see Thms.\ 
\ref{thm:well.behaved.from.smaller} and \ref{thm:well.behaved}.  The
ultimate results, including the factorization theorem will follow
quite easily---see Thms.\ \ref{thm:TGamma.accuracy},
\ref{thm:remainder.power.suppressed} and \ref{thm:asy.good.estimate},
and Secs.\ \ref{sec:factorization.overall} and
\ref{sec:cross.section}, etc. 

For our proofs to work, we have to consider non-leading regions as
well as leading regions, and the above definition also applies to
them.  By definition, to say a region is non-leading means that there
is a power suppression near the defining surface of the region.  There
is no need of subtractions for non-leading regions, of course.

The restriction to $M_{\rm ext}(k,\Gamma) < M_B(k,\Gamma)$ is needed, because
once the external momenta get further off-shell than is allowed by
this restriction, it becomes unnecessary to ask how close the internal
momenta $k$ are to the singular surface; the external virtuality has
already forced some of the internal lines to have at least comparable
virtuality.

The next theorem says that if all the leading singular regions of $\Gamma$
have been properly subtracted, then a good approximation to the subtracted
graph is obtained by applying a Taylor expansion to the dependence on its
external momenta:
\begin{theorem}
\label{thm:TGamma.accuracy}
  If $\Gamma$ is well behaved and $M_{\rm ext}(k,\Gamma) < M_B(k,\Gamma)$, then 
  \begin{equation}
  \label{eq:TGamma.accuracy}
    (1-V_\Gamma T) \overline\Gamma
    = O\!\left(
        L(\Gamma) \frac{ M_{\rm ext}^2(k,\Gamma) }{ Q^2 }
      \right)
  \end{equation}
  The left-hand-side is one of the equivalent forms of the definition
  of $\mathop{\rm Rem}\Gamma$---see Eq.\ (\ref{eq:def.Rem})---so this
  theorem is just stating that the remainder is power-suppressed (or,
  as is often said, it is of higher twist).
\end{theorem}
\begin{proof}
   $T$ replaces the external momenta of $\overline\Gamma$ by on-shell massless
   momenta and replaces the internal masses $m$ by zero.  In addition it
   inserts a Jacobian factor for the transformation between massive and
   massless momenta.  We now apply Thm.\ \ref{thm:accuracy.2} to $\Gamma$ when
   the region $R$ is the purely ultra-violet region for $\Gamma$, to obtain:
   \begin{equation}
   \label{eq:well.behaved.proof.1}
       \mathop{\rm Rem}\Gamma
    =
         (1-V_\Gamma T) \overline\Gamma
    = O\!\left(
        \| \overline\Gamma \|
        \frac{ M_{\rm ext}^2(k,\Gamma) }{ M_B^2(k,\Gamma) }
      \right)
   \end{equation}
   Here, we have used the fact that the distance to the defining
   surface of the ultra-violet region is of order $M_{\rm ext}$.  The
   desired Eq.\ (\ref{eq:TGamma.accuracy}) then follows from the
   meaning, Eq.\ (\ref{eq:well.behaved}), we gave to $\Gamma$ being
   well-behaved.

   More intuitively, this proof can be summarized as
   \begin{itemize}
   \item Away from all singular regions, the definition of $T$ implies
     that a graph minus its Taylor expansion is suppressed by a factor of
     order $M_{\rm ext}^2/Q^2$, which is already the desired result, Eq.\
     (\ref{eq:TGamma.accuracy}). 
   \item As each singular region is approached, this suppression
     is worsened by a factor $Q^2/\lambda^2(k,R)$.
   \item But because the hard subgraph of each region is well-behaved, by
     the hypothesis of the theorem, $\overline\Gamma$ is itself suppressed by a
     factor $\lambda^2(k,R)/Q^2$ relative to $\Gamma$.
   \item The last two factors cancel each other.
   \end{itemize}
   Consideration of the worse case scenario for each value of $k$
   requires us to replace $\lambda^2$ by $M_B^2(k,\Gamma)$.
   
   Although this proof looks quite trivial, there is actually one point
   that is particularly non-trivial.  This is in the application
   of Thm.\ \ref{thm:accuracy.2}.  That theorem applies most obviously to
   a graph or a function that is a simple product of propagators.  But
   when the graph is a difference of relatively large quantities, the
   derivation requires more care.
   
   A simple mathematical example for these results is provided by
   setting
   \begin{equation}
      \Gamma = \frac{ 1 }{ \left( m^2 + M_{12}^2 \right)
                      \left( m^2 + 2 M_{12}^2 +Q^2 \right)
                    },
   \end{equation}
   where $M_{12}$ is as an analog to a jet mass.  The subtraction of the
   singular region $M_{12} \ll Q$ is accomplished by writing
   \begin{eqnarray}
   \nonumber
      \overline \Gamma &=& \Gamma - T_1 \Gamma
   \\
   &=&
      \frac{ 1 }
          { 
             \left( m^2 + M_{12}^2 \right)
             \left( m^2 + 2 M_{12}^2 +Q^2 \right)
          }
      - 
      \frac{ 1 }{ \left( m^2 + M_{12}^2 \right) ~ Q^2 } .
   \end{eqnarray}
   Here $T_1$ sets $m$ and $M_{12}$ to zero in the 
   $1/(m^2 + 2 M_{12}^2 +Q^2)$ factor.
   
   The analog to Eq.\ (\ref{eq:TGamma.accuracy}) uses an operation $T$
   applied to $\overline\Gamma$, and in this model it consists of setting $m$
   to zero:
   \begin{equation}
      T\overline\Gamma
   =
      \frac{ 1 }{ M_{12}^2 \left( 2 M_{12}^2 +Q^2 \right) }
      - 
      \frac{ 1 }{ M_{12}^2 Q^2 } .
   \end{equation}
   There are relative errors of order $m^2/M_{12}^2$
   in both terms separately.  At first sight we only have
   \begin{equation}
      (1-T) \overline\Gamma = 
       O\!\left( \frac{ m^2 }{ M_{12}^2 } \right)
       \times \Gamma
       - O\!\left( \frac{ m^2 }{ M_{12}^2 } \right)
       \times C_R\Gamma ,
   \end{equation}
   which gives $O(\Gamma m^2 / M_{12}^2)$.  Since $M_{12}$ may be much
   smaller than $Q$, this error may be much larger than the desired
   error, where the denominator in the error is $Q^2$. However, the
   biggest parts of the errors are in fact identical, and this gives
   the correct result $O(\Gamma m^2 / Q^2)$.
   
   The general approach to this issue is to combine all the terms in
   $\overline\Gamma$ so that they are over a common denominator.  Then the
   approximation is applied to each factor in turn, and hence the error is
   the relative error produced by the $1-T$ operation times the size of
   $\overline\Gamma$.
\end{proof}

\begin{theorem}
\label{thm:well.behaved.from.smaller}
  Suppose all graphs smaller than $\Gamma$ are well-behaved, in the sense
  of Def.\ \ref{def:well.behaved}, so that $\Gamma$ is itself well-behaved.
\end{theorem}
\begin{proof}
  The structure of the proof generalizes the treatment of the concrete
  examples in Sec.\ \ref{sec:examples}, and the proof will be
  illustrated with the aid of these examples.

  If $\Gamma$ is small enough that it has no singular regions,
  then $M_B=Q$ and $\overline\Gamma = \Gamma$.  Then the definition of
  well-behavedness is trivially satisfied.
  
  If $\Gamma$ does have singular region(s), let $k$ be any point for $\Gamma$
  that satisfies $M_{\rm ext}(k,\Gamma) < M_B(k,\Gamma)$, and let $R$ be the
  {\em nearest} singular surface.  We put the other singular
  surfaces $R'$ into the following classes:
  \begin{itemize}
  \item[(a)] $R'<R$,
  \item[(b)] $R'>R$,
  \item[(c)] $R'$ intersects $R$, but it is neither a subset nor a
    superset of $R$, 
  \item[(d)] $R'$ does not intersect $R$ at all: $R \cap R' = \emptyset$.
  \end{itemize}
  Then
  \begin{eqnarray}
  \label{eq:Gammabar.split}
  \nonumber
    \overline\Gamma
    &=& \Gamma - \sum_{R'<R} C_{R'}\Gamma - C_{R}\Gamma 
       - \sum_{R'\text{ in (b--d)}} C_{R'}\Gamma 
  \\
    &=& (1-V_RT_R) \overline C_{R}\Gamma 
       - \sum_{R'\text{ in (b--d)}} C_{R'}\Gamma ,
  \end{eqnarray}
  where we used Eqs.\ (\ref{eq:CRbar.def}) and (\ref{eq:CR.from.CRbar}).
  We need estimates for each of these terms.  Note first that if $R$ is a
  {\em non}-leading region, then the subtraction term $C_{R}\Gamma$ for
  this region is zero, and thus that $T_R$ is zero for a non-leading
  region.  The necessary estimates for the terms corresponding to each
  class of $R'$ are obtained as follows:
  \begin{itemize}
  \item[(a)] From Eq.\ (\ref{eq:CRbar.Hbar}),
    \begin{equation}
       (1-V_RT_R) \overline C_{R}\Gamma 
       = \left( \mbox{jets of $R$} \right)
       (1-V_RT_R) \overline{H}_R ,
    \end{equation}
    where $\overline{H}_R$ is the subtracted hard subgraph of region $R$.
    Since 
    $H_R$ is a smaller graph than $\Gamma$, it is well-behaved, by the
    hypothesis of the theorem.  Moreover, because $R$ is the closest
    singular surface to $k$, the propagators with lowest
    virtuality are in its jet (or soft) subgraphs, and so 
    $\lambda(k,R) \lesssim M_{\rm ext}(k,H) \lesssim M_B(k,H)$. 
    Hence Thm.\ \ref{thm:TGamma.accuracy} gives
    \begin{equation}
       (1-V_RT) \overline{H}_R
       = O\!\left(
            L(H_R) \frac{ \lambda^2(k,R) }{ Q^2 }
         \right) ,
    \end{equation}
    and so
    \begin{eqnarray}
    \label{eq:term.a}
    \nonumber
       (1-V_RT_R) \overline C_{R}\Gamma 
       &=& O\!\left(
            L(\Gamma) \frac{ \lambda^2(k,R) }{ Q^2 }
         \right) 
    \\
       &=& O\!\left(
            L(\Gamma) \frac{ M_B^2(k,\Gamma) }{ Q^2 }
         \right) ,
    \end{eqnarray}
    where the last statement follows since $R$ is the closest singular 
    surface to $k$, and the definition of $M_B(k,\Gamma)$ is that it equals
    the distance to the closest singular surface.
    
    An example of this result is given by setting $R=R_2$ for Fig.\ 
    \ref{fig:4jet}.  In this case, Eq.\ (\ref{eq:term.a}) states that
    $\Gamma - \Gamma_1 - \Gamma_2 = O(\Gamma M_{34}^2/Q^2)$---as we saw at Eq.\ 
    (\ref{eq:R2.decomp}).

  \item[(b)] Consider a singular region $R'>R$.  We only need consider
    a leading region for $R'$, since otherwise $C_{R'}\Gamma = 0$. The
    definition of $C_{R'}$ gives 
    \begin{eqnarray}
    \nonumber
       C_{R'}\Gamma 
       &=& V_{R'}T_{R'} C_{R'}\Gamma 
    \\
       &=& \left( \mbox{ jets of $R'$ } \right)
           V_{R'}T_{R'} \overline{H}_{R'} .
    \end{eqnarray}
    Since $R'>R$ the hard subgraph of $R'$ is a {\bf bigger} graph than
    the hard subgraph of $R$, and some of the lines of the virtuality that
    determines $R$ as the closest singular surface to $k$ must lie in
    this hard subgraph of $R'$.
    
    Moreover $R'$ is a non-trivial singular surface, so its hard
    subgraph is smaller than $\Gamma$, so it is well-behaved, by the
    theorem's hypothesis.  Since $T_{R'} \overline{H}_{R'}$ has its external
    lines on-shell, we can apply the definition of well-behavedness to
    give 
    \begin{eqnarray}
       T_{R'} \overline{H}_{R'}
       &=& O\!\left(
            L(H_{R'}) \frac{ \lambda^2(k,R) }{ Q^2 }
         \right) 
    \nonumber\\
       &=& O\!\left(
            L(H_{R'}) \frac{ M_B^2(k,\Gamma) }{ Q^2 }
         \right) ,
    \end{eqnarray}
    from which follows
    \begin{equation}
    \label{eq:term.b}
       C_{R'} \Gamma
       = O\!\left(
            L(\Gamma) \frac{ M_B^2(k,R) }{ Q^2 }
         \right) .
    \end{equation}
    An example is given by setting $R=R_1$ and $R'=R_2$ for Fig.\ 
    \ref{fig:4jet}.  Then Eq.\ (\ref{eq:term.b}) states that $\Gamma_2 =
    O[\Gamma(M_{12}^2+M_{34}^2)/Q^2]$.

  \item[(c)] Suppose $R'$ non-trivially intersects $R$.  Just as in the
    previous case, we have
    \begin{equation}
    \label{eq:term.c}
       C_{R'}\Gamma 
       = \left( \mbox{ jets of $R'$ } \right)  V_{R'}T_{R'} \overline{H}_{R'} .
    \end{equation}
    We are considering a singular surface that is closest to
    $k$.  In this region a certain set of lines has low virtuality.
    Now in the factor $T_{R'} \overline{H}_{R'}$, a certain other set of lines
    is set on-shell, and effectively this factor is being calculated in a
    neighborhood of the intersection of the two regions, $R \cap R'$.
    This case reduces now to case $(b)$.
    
    For Fig.\ \ref{fig:4jet}, an example is given by setting $R=R_3$
    and $R'=R_2$, so that $R \cap R' = R_1$, and Eq.\ (\ref{eq:term.c})
    becomes $\Gamma_2 = O(\Gamma M_{12}^2/Q^2)$---see Eq.\ 
    (\ref{eq:special.suppression}).
    
  \item[(d)] The final possibility is that $R' \cap R = \emptyset$.  The
    definition of $R$ implies that $k$ is close to $R$.  Now, in
    $C_{R'}\Gamma$, the hard subgraph $\overline{H}_{R'}$ has its external
    lines on shell, so $k$ being near to $R$ implies also that in the
    hard subgraph the momentum configuration is close to $R' \cap R$.
    But there are no such points.  

  \end{itemize}
  We have shown that all the terms on the right-hand-side of Eq.\
  (\ref{eq:Gammabar.split}) are of order $L(\Gamma) M_B^2(k,\Gamma)/Q^2$,
  and the theorem follows.
\end{proof}

\begin{theorem}
\label{thm:well.behaved}
  All graphs $\Gamma$ are well-behaved.
\end{theorem}
\begin{proof}
  This is a simple consequence of mathematical induction applied to
  the previous theorem \ref{thm:well.behaved.from.smaller}.  If there
  are graphs $\Gamma$ that are not well-behaved, then there is a minimal
  such graph.  Application of Thm.\ 
  \ref{thm:well.behaved.from.smaller} shows that the graph is
  well-behaved, contrary to hypothesis.
\end{proof}

\begin{theorem}
\label{thm:remainder.power.suppressed}
  The remainder is power-suppressed for an on-shell graph:
  \begin{equation}
     \mathop{\rm Rem}\Gamma 
       = O\!\left(
            L(\Gamma) \frac{ m^2 }{ Q^2 }
         \right) .
  \end{equation}
\end{theorem}
\begin{proof}
  From the definitions, $\mathop{\rm Rem}\Gamma = (1-T)\overline\Gamma$, and
  from the previous theorem, $\Gamma$ is well-behaved.  Hence by Thm.\
  \ref{thm:TGamma.accuracy} 
  \begin{equation}
     \mathop{\rm Rem}\Gamma 
       = O\!\left(
            L(\Gamma) \frac{ M_{\rm ext}^2(k,\Gamma) }{ Q^2 }
         \right) 
       = O\!\left(
            L(\Gamma) \frac{ m^2 }{ Q^2 }
         \right) .
  \end{equation}
\end{proof}

It immediately follows that
\begin{theorem}
\label{thm:asy.good.estimate}
  $\mathop{\rm Asy}\Gamma$ is a good estimate of $\Gamma$ up to
  power-law corrections, i.e.,
  \begin{eqnarray}
     \Gamma &=& \mathop{\rm Asy} \Gamma 
              + O\!\left( L(\Gamma) \frac{m^2}{Q^2} \right)
  \nonumber\\
      &=& \sum_R C_R\Gamma 
              + O\!\left( L(\Gamma) \frac{m^2}{Q^2} \right) .
  \end{eqnarray}
\end{theorem}

\begin{theorem}
\label{thm:IR.safety.H}
  On-shell subtracted hard subgraphs $T\overline{H}_R$ are IR safe.  That is,
  in the integral over final-state momenta, no collinear singularities
  result when two or more of the external lines of a hard subgraph become
  collinear.
\end{theorem}
\begin{proof}
  This theorem is, of course, essential for effective perturbative
  calculations.  Because the masses of internal and external lines are set
  to zero by the $T$ operation, the logarithmic enhancements in the
  collinear regions for the unsubtracted graphs become actual logarithmic
  divergences after the integral over final states.
  
  The IR safety follows from the theorem that any graph, e.g., $H_R$, is
  well-behaved: for then, in the neighborhood of any of the singularities,
  $\overline{H}_R$ has a power-law suppression compared with the unsubtracted
  graph.
\end{proof}

\subsection{Factorization for the cross section}
\label{sec:cross.section}

We now have all the elements for use in the skeleton of the factorization
proof in Sec.\ \ref{sec:proof.summary}: a definition of $C_R(\Gamma)$ and a
proof that the remainder in Eq.\ (\ref{eq:sum.over.regions}) is indeed
power suppressed.  The factorization theorem Eq.\ 
(\ref{eq:basic.factorization}) has therefore been proved.

\begin{figure}
   \includegraphics{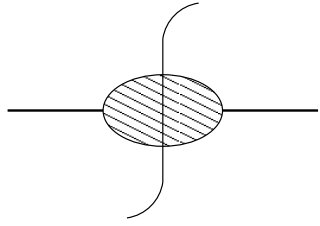}
   \caption{Jet factor}
   \label{fig:jet.factor}
\end{figure}

In Eq.\ (\ref{eq:factorization2}), we expressed the factorization formula
in terms of the phase-space integral for massless partons.  Now we will
reorganize this formula into a form that is particularly convenient for
calculation, and especially for calculations by a MC event generator.  It
will consist of a hard-scattering cross section for making $N$ partons
multiplied by the probabilities for each of the partons to hadronize to a
particular state.  The sum over the (conditional) hadronization
probabilities is of course to be unity,

To this end, we first define an integrated jet factor
\begin{equation}
\label{eq:Z.definition}
  Z(\mu_F/\mu_R,m/\mu_R) = 
   \int \frac {dM_j^2}{2\pi }
   \int dL(p_j;l_j) \Theta(l_j^2/\mu_F^2)  J_j(p_j) .
\end{equation}
As before, $J_i$ is the sum over graphs of the form of Fig.\ 
\ref{fig:jet.factor}, and we integrate it over all possible final
states with the cut-off function $\Theta(l_j^2/\mu_F^2) = \Theta(M_j^2/\mu_F^2)$
restricting the range of allowed jet masses.  This integral is
performed in the rest-frame of the final-state of the jet, i.e., we
set $l_j^\mu = (M_j, \JCCvec{0})$.  The quantities $\mu_F$ and $\mu_R$ are
the factorization and renormalization scales, and it is the cut-off
function gives the factorization scale its meaning, since nothing else
in Eq.\ (\ref{eq:Z.definition}) depends on $\mu_F$.

Next we define a differential jet factor as
\begin{equation}
\label{eq:jet.factor.def}
   dD(p_j; l_j; \mu_F) = 
           dL(p_j;l_j,m) ~\Theta(M_j^2/\mu_F^2) 
           ~ \frac{ J_j(p_j) }{ Z(\mu_F/\mu_R) } .
\end{equation}
This is the sum over fragmentation graphs completely differential in
the phase space for the final-state particles.  Note carefully that
the phase-space measure $dL$ contains the exact jet momentum $l_j$,
not the approximated jet momentum $\hat l_j$.  To allow a probability
interpretation, a normalization factor $1/Z$ is present so that the
integral of the factor over final states and masses is unity:
\begin{equation}
\label{eq:jet.sum}
   \int \frac{ dM_j^2 }{ 2\pi } \int dD(p_j;l_j) = 1 .
\end{equation}
(This integral is to be performed in the jet's rest frame.)

The hard scattering cross section is defined by
\begin{equation}
  \label{eq:sigma.hat}
  d\hat{\sigma}_N = K~ dL(\hat l;q,0) ~ \hat{H}_N ~ Z(\mu_F)^N .
\end{equation}
This is the sum of fully subtracted hard scattering graphs times a
factor $Z(\mu_F)$ for each final-state parton.  We have included the
factor $K$, and a phase-space measure so that $d\hat\sigma$ really does
have the normalization of a parton-level differential cross section.
Observe that $Z$ plays the role of a wave-function renormalization
factor, just like the factor that would be obtained from the usual LSZ
reduction method for a cross section.  But note carefully that the
correct procedure for calculating the hard-scattering cross section
$\hat\sigma$ is to use the integrated jet factor $Z(\mu_F)$ in place of the
residue of the propagator pole that would be indicated by the LSZ
theorem. 

With these definitions, the factorization formula Eq.\ 
(\ref{eq:factorization2}) can be rewritten as
\begin{equation}
\label{eq:factorization3}
   \sigma[W]
   =
   \sum_N
   \int d\hat\sigma_N(\hat l)
   \times \prod_{j=1}^{N} \int \frac{ dM_j^2 }{ 2\pi } 
   \times \prod_{j=1}^{N} \int dD(p_j;l_j) 
   \times W(f)
   +  \mbox{non-leading power} .
\end{equation}
In the last line we have a hard-scattering cross section $\hat{\sigma}_N$ and
jet factors $dD(p_j)$. The integral over the jet momenta is expressed in
terms of the massless momenta $\hat l$, since we will calculate the
hard-scattering factor from massless on-shell graphs (with the aid of some
subtractions); this results in the disappearance of the Jacobian $\Delta$
compared with Eq.\ (\ref{eq:basic.factorization}).

The power-law correction in Eq.\ (\ref{eq:factorization3}) is of
order the cross section times $m^2/Q^2$, with the usual addition of a term
associated with non-leading regions.

\subsection{Infra-red finiteness of integrated jet factor $Z(\mu_F)$}
\label{sec:IR.safety.int.jet}

\begin{figure}
\psfrag{mu2}{$\mu_F^2$}
\psfrag{Re}{${\rm Re}(M^2)$}
\psfrag{Im}{${\rm Im}(M^2)$}
\includegraphics{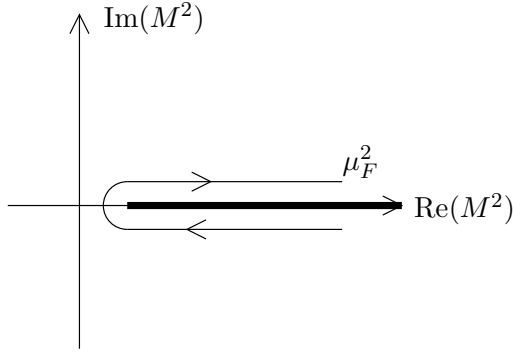}
\caption{Contour used to define the integrated jet factor $Z(\mu_F)$.}
\label{fig:contour.1}
\end{figure}

The integrated jet factor $Z(\mu_F/\mu_R,m/\mu_R)$ defined by Eq.\ 
(\ref{eq:Z.definition}) is the integral of a cut propagator, Fig.\ 
\ref{fig:jet.factor}.  This implies that, even though the integrand
includes small jet masses, the integral is infra-red safe for large
$\mu_F$, so that it is a good approximation to set particle masses to
zero.  The proof is the same as for an integral over the hadronic part
of $e^+e^-$ cross section and for other similar quantities \cite{pc}.
It goes as follows: The integrated jet factor is the integral over the
discontinuity of the uncut propagator divided by $2\pi$:
\begin{equation}
   Z(\mu_F) = \int_{C_1} \frac{dM^2}{2\pi} \Theta(M^2/\mu_F^2) S(M^2), 
\end{equation}
where the contour $C_1$ encloses the cut of the propagator $S(M^2)$, as in
Fig.\ \ref{fig:contour.1}.  The contour can be deformed into the complex
plane to a region where the (complex) virtuality is of order $\mu_F^2$. Since
the internal virtualities are now large, setting the particle mass $m$ to
zero is a good approximation that is valid up to a power-law correction.

In the above derivation, we have assumed that the cut-off function is a
theta function $\Theta(M^2/\mu_F^2) = \theta(\mu_F-M)$.  If a smoother
function is used, minor modifications in the above demonstration are
needed.

\subsection{Renormalization group}
\label{sec:RG}

Let us use the term ``factorization scale'' to denote the scale $\mu_F$
that appears in the cut-off function $\Theta(m^2/\mu_F^2)$ in the above
formulae, and let the usual renormalization scale be $\mu_R$.

The renormalization-group operator is, as usual,
\begin{equation}
  \label{eq:RGop}
  \mu_R^2\frac{d}{d\mu_R^2} 
  = \mu_R^2\frac{\partial}{\partial\mu_R^2} 
    + \beta(g)\frac{\partial}{\partial g}
    - \gamma_m m^2 \frac{\partial}{\partial m^2} ,
\end{equation}
where $\gamma_m$ is the anomalous dimension of the mass squared, and
\begin{equation}
  \label{eq:beta}
  \beta(g) = - \frac{ 3g^2 }{ 512\pi^3 }
         + O(g^5) .
\end{equation}
Then the integrated jet factor obeys
\begin{equation}
  \label{eq:RG.Z}
  \mu_R^2\frac{dZ(\mu_F/\mu_R;g(\mu_R))}{d\mu_R^2}
  = -\gamma(g) Z(\mu_F/\mu_R) ,
\end{equation}
where the anomalous dimension of the parton field is
\begin{equation}
  \label{eq:gamma}
  \gamma(g) = \frac{ g^2 }{ 768\pi^3 } + O(g^4) .
\end{equation}

In QCD, the electromagnetic current is conserved and therefore has no
anomalous dimension.  It follows that the hard scattering factor, with the
appropriate factors of $Z(\mu_F)$ is renormalization-group-invariant.
However, in our model field the photon-parton coupling does not involve a
conserved current and it in fact has an anomalous dimension.  Thus the
renormalization group equation for the strong-interaction part of the hard
scattering coefficient does have an anomalous dimension:
\begin{equation}
  \label{eq:RG.sigma.hat}
  \mu_R\frac{d(\hat\sigma_N/K)}{d\mu_R}
  = \mu_R\frac{d(\hat{H}_N)Z^N}{d\mu_R}
  = -\beta_e(g) \hat{H}_NZ^N 
  = -\beta_e(g) \hat\sigma_N/K ,
\end{equation}
where $\beta_e$ is an anomalous dimension associated with the electromagnetic
coupling:
\begin{equation}
  \label{eq:beta.e}
  \beta_e(g) = - \frac{ 5g^2 }{ 768\pi^3 } .
\end{equation}


%% file: phi3-algorithm.tex

\section{Monte-Carlo algorithm}
\label{sec:monte.carlo.algorithm}

\subsection{Probabilities}

To obtain a Monte-Carlo algorithm, we observe that when the weight factor
$W(f)$ is omitted, the integrand of Eq.\ (\ref{eq:factorization3}) gives
the probability density of the final states.  The weight factor $W$ simply
gave a convenient way of unifying the derivations of cross sections into a
derivation of a total weighted cross section.  Therefore we can obtain any
cross section by the following technique: (a) omit the weighting factor
$W(f)$, (b) multiply by the integrated luminosity, and (c) use a
Monte-Carlo method to integrate and sum over the internal partonic
variables ($N$, $\hat l$, and $M_j$) and over the hadronic final states.
Recording the events generated by the Monte-Carlo integration, together
with their weights allows us to construct any binned cross section.

Because of the factorization, the density for the final state
is a product\footnote{
\label{foot:limit}
   One annoying problem is that the cut-off on jet masses must be less 
   than $Q/N$, for otherwise one can generate events that violate
   energy conservation.  There are at least two solutions: 
   (a) Set the cut-off $\mu_F$ to be smaller than $Q/N_{\rm max}$, where
   $N_{\rm max}$ is the largest number of partons in the finite-order
   hard-scattering calculations that one actually uses.  Then
   perform the somewhat complicated renormalization-group
   calculation that would enable the cut-off to be different for
   hard-scattering coefficients with different numbers of final-state
   partons.
   (b) Permit a higher cut-off, but veto events whenever they are
   generated with $\sum_j M_j > Q$.  For consistency, a corresponding veto
   will need to be built into the subtraction terms in hard-scattering
   calculations.  Note that the veto will only affect events in which
   a least one of the generated jets has an invariant mass of order
   $Q$. 
}
of 
\begin{itemize}
\item A density for particular partonic final states (variables
  $N$ and $\hat{l}$).  This is given by the differential partonic cross
  section Eq.\ (\ref{eq:sigma.hat}).
\item A (conditional) density for each jet mass $M_j$, given
  the partonic variables.
\item A (conditional) density for the final-state for each
  jet, given its mass $M_j$.
\end{itemize}
Aside from the issue of negative weight terms, the densities are to be
interpreted as probability densities, that is they sum and integrate to
unity. 

For determining the jet mass, it is convenient to define the integral over
the jet factor $J_i$ over its final states.  With a normalization factor
that will prove convenient, it is
\begin{equation}
  \label{eq:F}
  F(M_j; \mu_R, m(\mu_R), g(\mu_R))
  = \frac{M_j^2}{2\pi} \int dL(p_j;l_j) 
    J_j(p_j; l_j; \mu_R, m(\mu_R), g(\mu_R)) .
\end{equation}
(The integral over the hadronic state is performed for a fixed value of
the total jet momentum $l_j^\mu$.)
Apart from the normalization factor, this is just the cut propagator $S$,
and it is simply related to the integrated jet factor:
\begin{equation}
   F(M_j) = \frac{M_j^2~S(M_j^2)}{2\pi} 
        = M_j^2 \left. \frac{\partial Z(\mu_F)}
                            {\partial \mu_F^2} \right|_{\mu_F=M_j} .
\end{equation}
Hence the massless limit can be taken for large jet masses: $F$ is
infra-red safe.  Notice that $F$ depends on the renormalization scale, but
not on the factorization scale.  Since it is proportional to a cut
propagator, it has anomalous dimension $\gamma$:
\begin{equation}
  \label{eq:F.RGE}
  \mu_R^2 \frac{ dF }{ d\mu_R^2 } = -\gamma(g(\mu_R)) ~ F .
\end{equation}
Then the distribution of jet masses is given by
\begin{equation}
\label{eq:dP.dM2.0}
      \frac{ dP(M_j^2; \mu_F) }{ dM_j^2 }
      = \frac{1}{M_j^2} 
        \frac{ F(M_j^2; \mu_R) ~ \Theta(M_j^2/\mu_F^2)  }{ Z(\mu_F/\mu_R) } .
\end{equation}
This is renormalization-group invariant, since it is the quotient of two
quantities with the same anomalous dimension.

If we set $\mu_F$ and $\mu_R$ to be of order $Q$, then a perturbative
calculation of the hard scattering is valid.  However the calculation of
probability density for $M_j$ suffers from large logarithms whenever $M_j
\ll \mu_R$, so we must perform a renormalization group improvement for the
calculation of the $M_j^2$ distribution.  This enables $F$ to be
calculated with a renormalization scale $M_j$ (a different scale for each
jet):
\begin{equation}
\label{eq:dP.dM2}
   \frac{ dP(M_j^2; \mu_F) }{ dM_j^2 }
      = \frac{1}{M_j^2}
        \frac{ F(M_j^2, M_j, g(M_j) ) ~ \Theta(M_j^2/\mu_F^2)  }
             { Z(\mu_F/\mu_R) }  
        \exp\left[ - \int_{M_j^2}^{\mu_R^2} \frac{d\mu'^2}{\mu'^2} 
                   \gamma(\alpha_s(\mu'))
             \right] .
\end{equation}
{\em Observe that the renormalization scale $\mu_R$ has been replaced by
$M_j$ in the numerator $F$ but not in the denominator $Z$.}  Then a
finite-order calculation of $F$ is valid provided only that $M_j \gg
\Lambda$, so that $g(M_j)$ is small.  Of course, if $M_j$ is small, one
must resort to phenomenological functions and models that are adjusted to
fit data, just as in current event generators.  I have set the
renormalization scale to $M_j$ for the jet, but any value of similar size
would be equally good in principle.  A somewhat smaller size would
probably be best if the \MSbar{} scheme is used, since the typical
virtualities in the higher order calculation will be smaller than $M_j^2$.

This formula is very similar to the formulae used in standard MC event
generators.  For example, the integral over the anomalous dimension
corresponds exactly to the Sudakov form factor in the standard LLA
formulation \cite{MC.Review}, with the anomalous dimension being an
integral of the DGLAP evolution kernel over the splitting variable $z$.
To leading order, the $1/Z$ factor can be replaced by its lowest order
term, unity.  Also, it can be verified that at leading (one-loop) order
$F$ is the anomalous dimension of the parton field:
\begin{equation}
  \label{eq:F.from.gamma}
  F = \gamma(g) + O(g^4) .
\end{equation}
Thus the leading approximation to the distribution of $M_j^2$ is
\begin{equation}
\label{eq:dP.dM2.leading}
   \frac{ dP(M_j^2; \mu_F) }{ dM_j^2 }
      = \frac{1}{M_j^2} \gamma(g(M_j)) ~ \Theta(M_j^2/\mu_F^2)
        \exp\left[ - \int_{M_j^2}^{\mu_R^2} \frac{d\mu'^2}{\mu'^2} 
                   \gamma(g(\mu'))
             \right] 
        \left( 1 + O(g(M_j)^2) \right).
\end{equation}

However, if one goes beyond leading order some differences become
apparent.  The $Z$ factor is no longer unity, and $F$ is not the same as
the anomalous dimension; both of these quantities are nevertheless
perturbatively calculable (if $\mu_F$ and $M_j$ are large).  

So far we have discussed the hard scattering probabilities and the
calculation of the values of $M_j$ for each jet.  The final ingredient
is the distribution of the final states of a particular jet given
$M_j$:
\begin{equation}
\label{eq:dJ}
    dJ = \frac{ dL(p_j;l_j) J_j(p_j,M_j;\mu_R:=M_j) }{ F(M_j; \mu_R:=M_j) } .
\end{equation}
The denominator factor ensures that the integral over all final states is
unity.  In addition, the numerator and denominator have the same anomalous
dimension, so that $dJ$ is renormalization-group invariant.  We may
therefore set the renormalization scale to $M_j$, as indicated.  Since the
denominator is an integrated quantity with external mass $M_j$, it can now
be calculated perturbatively without large logarithms.  The numerator has
an external mass $M_j$ with an equal renormalization scale.

Because the numerator is exclusive, a fixed-order perturbative calculation
suffers from large logarithms.  However it is an object of the same type
as the cross section that we originally tried to calculate: The problem of
generating the final state from the fragmentation of the $j$th parton
given its invariant mass $M_j$ is essentially the same as the calculation
of the final state of the whole event given $Q$.  So we must recursively
apply our algorithm to each jet.

To this end we need a factorization formula for the jet, and the only
difference between this formula and the corresponding formula for the
cross section is that the hard-scattering cross sections $d\hat\sigma_N$ are
replaced by evolution kernels $d\hat J_N$:
\begin{equation}
\label{eq:jet.factorization}
  dJ = \sum_N d\hat J_N \times \mbox{Jet factors} ,
\end{equation}
where the jet factor is the same as in Eq.\
(\ref{eq:factorization3}). Note the sum rule:
\begin{equation}
  \int \sum_N d\hat J_N = 1.
\end{equation}

\subsection{Algorithm}

Since the number of variables to specify the final state is large, the use
of Monte-Carlo techniques for the integration is appropriate.  We have
formulated the cross section as a product of factors, so that the
following algorithm generates events with the correct probability
distribution:
\begin{enumerate}

\item
    Generate a partonic final state according to the distributions given
    by the hard-scattering cross sections $d\hat\sigma_N$ defined in Eq.\
    (\ref{eq:sigma.hat}). Let the state contain $N$ on-shell partons, and
    let their momenta be $\hat l_j$ (for $j = 1, \dots N$).  If the
    hard-scattering cross sections are positive, then unweighted events
    can be generated, with the hard-scattering cross sections being
    treated as determining a probability distribution.  If the
    hard-scattering cross sections are negative in some region, then
    appropriately weighted events must be generated.

\item
    For each parton $j$
    \begin{enumerate}
    \item
        Choose a value $M_j$ for the resulting jet, according to the
        distribution in Eq.\ (\ref{eq:dP.dM2}).
    \item
        In the rest frame of the jet, construct the jet by a call to
        a showering algorithm.  This starts at step 1, using the
        probability density $d\hat J_N$ in Eq.\
        (\ref{eq:jet.factorization}) instead of the probability density
        that corresponds to $d\hat\sigma_N$.
    \item
        In its rest frame, let the total momentum of the jet be
        ${l'_j}^\mu = (M_j,\JCCvec{0})$, and let the final-state hadron
        momenta be $p'_{j,i}$.
    \end{enumerate}

\item \label{renormalize.momenta}
    After all the jets have been constructed, apply a boost to each
    jet in the direction of its parent parton.  Arrange the boost so
    that the total momentum of the jet in the overall center-of-mass
    frame obeys Eqs.\ (\ref{eq:lj.construction}) and
    (\ref{eq:lambda.def2}). 

\end{enumerate}
In this algorithm, the showering of each parton is performed in the rest
frame of the parton, with appropriate variables being the angles of the
resulting partons.  In contrast, many current event generators
\cite{MC.Review} use a momentum fraction variable, since the fragmentation
is conceived of as being in a frame in which the partons are fast moving.
The two views are necessarily equivalent.  

In order to have a solution of Eq.\ (\ref{eq:lambda.def2}) for $\lambda $, we
must require that $\sum_j M_j \leq Q$.  If we are to get results that
are independent of the order in which the jets are fragmented, this
implies that we most impose the condition $M_j \leq Q/N$ for each jet
separately.  This implies a maximum to the possible values of $\mu_F$.
Alternatively, as explained in footnote \ref{foot:limit}, we can apply
a veto whenever $\sum_j M_j>Q$, and insert a corresponding correction in
the subtraction terms in the hard-scattering coefficients.

As is usual in perturbative calculations we have a choice of
renormalization scale, to which is now added a choice of the cut-off
function $\Theta(M^2/\mu_F^2)$.  The choice should ideally made with the aim of
minimizing the impact of higher-order uncalculated perturbative
corrections.  The cross-section is invariant under changes of this choice
if the cross section and showering are calculated exactly.

But we have another choice in the algorithm.  This is the correspondence
between the approximated and unapproximated jet momenta at step
\ref{renormalize.momenta}.  Any change in this correspondence amounts to a
change of factorization scheme, and is compensated by corresponding
changes in the subtraction terms that are used in the hard scattering
cross section and in the splitting kernels. Such a choice always occurs in
a Monte-Carlo event generator, but the necessity to construct a systematic
algorithm for higher order corrections appears to me to restrict the range
of simple choices.

The definition I have used---Eqs.\ 
(\ref{eq:l.hat})--(\ref{eq:lambda.def2})---seems to be the simplest if one
requires that:
\begin{itemize}
  
\item Conservation of 4-momentum is obeyed at every stage.
    
\item The renormalization of momenta is symmetric in the jets, and in
  particular it should not depend on the order in which the jets are
  fragmented.
    
\item The fragmentation of a jet can be done independently of the
  momentum renormalization.  This implies that the momentum
  renormalization amounts to a Lorentz transformation of each jet, but
  a different transformation for each jet.
    
\item We must also allow for the possibility, in a more complicated
  theory than $\phi^3$, that there may be polarizations for the partons.
  In that case there are Lorentz spinors and vectors associated with
  the partons.  To avoid complicating the Lorentz transformations that
  implement step \ref{renormalize.momenta} of the algorithm, we can
  require that the Lorentz transformations be simple boosts, for which
  the only natural direction is $\JCCvec{l}_j$.

\end{itemize}


%% file: phi3-calculations.tex

\section{Examples of calculations}
\label{sec:calculation.examples}

In this section I summarize the results of calculations that give the
lowest-order quantities and the simpler next-to-lowest-order quantities.
I assume throughout that the cut-off function defining the factorization
scale provides a sharp cutoff:
\begin{equation}
  \Theta(M^2/\mu_F^2) = \theta(\mu_F-M) .
\end{equation}

\subsection{Integrated jet factor}

\begin{figure}
    \includegraphics[scale=0.6]{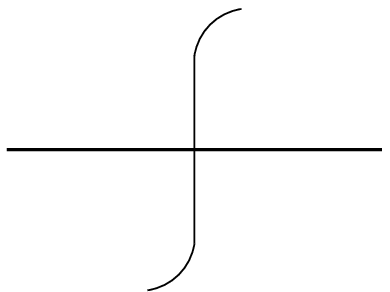}
    ~\includegraphics[scale=0.6]{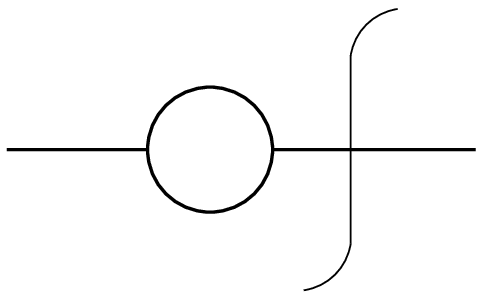}
    ~\includegraphics[scale=0.6]{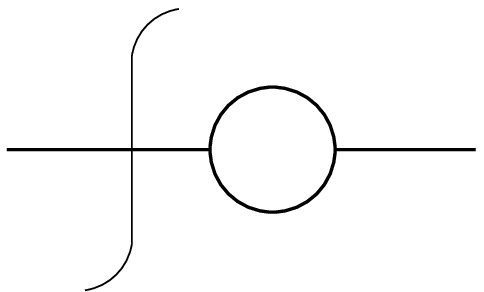}
    \\
    \includegraphics[scale=0.4]{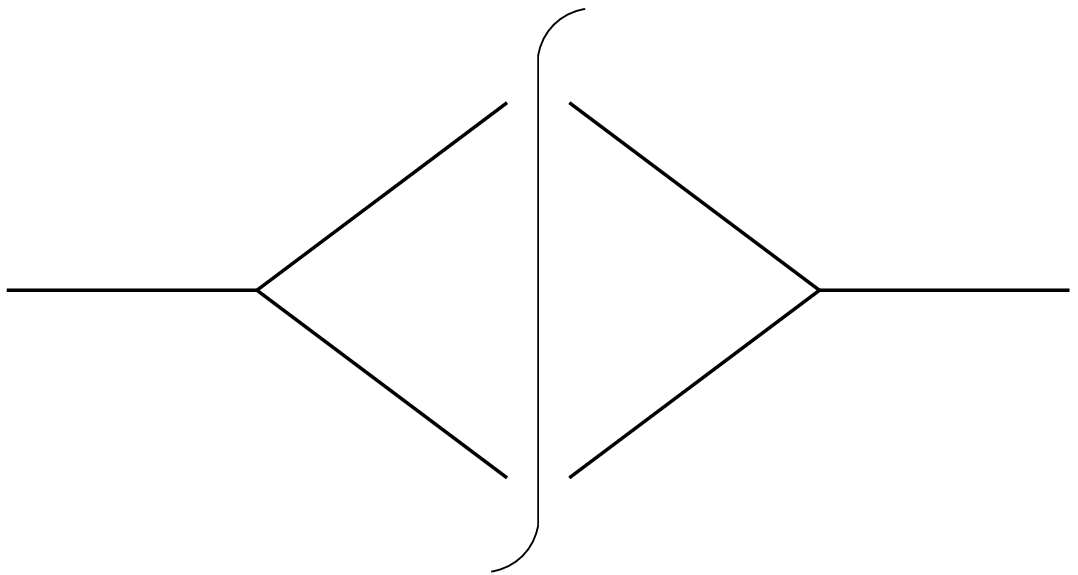}
\caption{Diagrams up to one-loop order for jet factor.}
\label{fig:jet-factor}
\end{figure}

Since the fully integrated jet factor is infra-red safe, it can be
calculated in massless limit from the graphs of Fig.\ \ref{fig:jet-factor}:
\begin{eqnarray}
  \label{eq:Z.calc}
  Z
  &=&
  \lim_{\epsilon\to0} \int dM_j^2 
  \left\{
    \left[ 1 + \frac{g^2 }{ 768\pi^3\epsilon }
    \right] \delta(M_j^2)
  + \frac{ g^2 }{ 768\pi^3 M_j^2 }
    \left( \frac{ 4 e^{\gamma_E} \mu_R^2 }{ M_j^2 } \right)^\epsilon
    \frac{ \Gamma(5/2) }{ \Gamma(5/2-\epsilon) }
  \right\}
  + O(g^4) 
\nonumber\\
  &=&
    1 + \frac{g^2 }{ 768\pi^3 }
        \left( \ln\frac{ \mu_F^2 }{ \mu_R^2 } - \frac{8}{3} \right) 
  + O(g^4) .
\end{eqnarray}
Here I used dimensional regularization with a space-time dimension
$6-2\epsilon$. The last factor in the braces arises from the integral over the
one-loop graph for real emission, and \MSbar{} renormalization has been
implemented by setting the bare coupling to be
\begin{equation}
  \label{eq:MSbar}
  g_0^2 = g_R^2 \left( \frac{ \mu_R^2 e^{\gamma_E} }{ 4\pi } \right)^\epsilon
            \left( 1 + \mbox{pure-pole counterterms}\right) ,
\end{equation}
where the counterterms are poles at $\epsilon=0$.

We also need the jet factor $F$ for non-zero jet mass.  In the massless
limit in the physical space-time dimension, it is 
\begin{equation}
  \label{eq:F.LO}
  F = \frac{ g^2 }{ 768\pi^3 } + O(g^4) .
\end{equation}
As stated earlier, this equals the anomalous dimension $\gamma$ at the leading
order.

\subsection{Distribution of partons in showering}

The square of the order $g$ graph for parton splitting gives a uniform
angular distribution in the rest frame of the parent parton.  Thus the
lowest order for the splitting probability is
\begin{equation}
  \label{eq:dJ2}
  d\hat{J}_2 = \frac{ d\Omega }{ 8\pi^2/3 } + O(g^4) ,
\end{equation}
where $d\Omega$ represents an element of solid angle in 5 space dimensions, and
the denominator is the total 5-dimensional solid angle.

\subsection{NLO hard-scattering coefficient}

The two-parton hard-scattering cross section---see Eq.\ 
(\ref{eq:sigma.hat})---is obtained from the lowest-order graph of Fig.\ 
\ref{fig:2jet}, from its one-loop virtual correction, and from multiplying
the result by the square of the jet factor $Z$.  The whole calculation is
done in the massless limit and the result is
\begin{equation}
  \label{eq:sigma.2}
  d\hat{\sigma}_2 =
  K dL(\hat{l}_1, \hat{l}_2 ;q, 0)
  \left[
    1 + \frac{ g^2 }{ 64\pi^3 }
        \left( -\ln\frac{Q^2}{\mu_R^2} 
               + \frac{1}{6}\ln\frac{\mu_F^2}{\mu_R^2}
               + \frac{14}{9}
        \right)
    + O(g^4)
  \right] ,
\end{equation}
where $K$ is the standard overall normalization factor for the cross
section. 

\begin{figure}
   \psfrag{p_1}{$\hat{l}_1$}
   \psfrag{p_2}{$\hat{l}_2$}
   \psfrag{p_3}{$\hat{l}_3$}
   \includegraphics[scale=0.6]{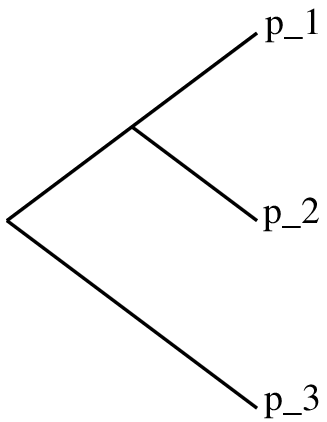}
   ~~\includegraphics[scale=0.6]{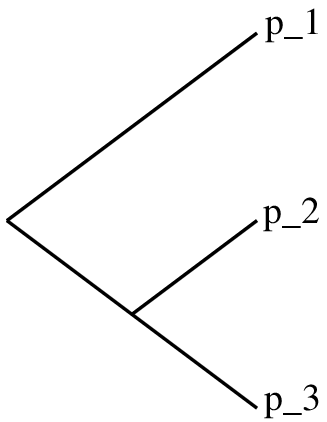}
   ~~\includegraphics[scale=0.6]{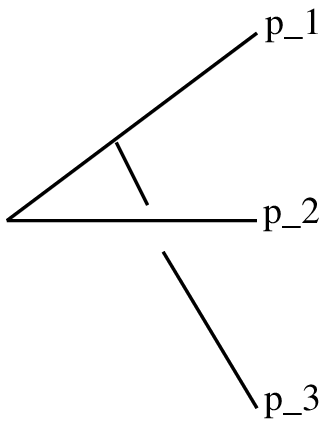}
\caption{Amplitude for three-parton production.}
\label{fig:3parton}
\end{figure}

The lowest-order amplitude for the three-parton cross section is given by
the sum of the three graphs in Fig.\ \ref{fig:3parton}.  Since we are
computing the hard-scattering coefficient, the external momenta are the
approximated, massless parton momenta, $\hat{l}_i$.  Before subtractions,
the squared amplitudes give
\begin{equation}
  \label{eq:sigma.3.unsub}
  g^2 \left( \frac{1}{M_{12}^2} + \frac{1}{M_{23}^2} + \frac{1}{M_{13}^2}
     \right)^2 .
\end{equation}
Here $M_{12}^2=(\hat{l}_1+\hat{l}_2)^2$, etc., are the masses of pairs of
partons.  When integrated over the parton kinematics, this squared
amplitude would give the well-known logarithmic collinear divergences.  

We have already calculated the subtractions for the square of one of the
graphs in Sec.\ \ref{sec:simple.example}.  There are two other terms,
obtained by permuting the labels for the final-state partons.  It is
convenient to use the conventional dimensionless variables for the parton
energies in the overall center-of-mass frame: $x_i=2E_i/Q=2\hat{l}_i^0/Q$.
These lie in the range $0\leq x_i\leq1$, and they obey $x_1+x_2+x_3=2$.
In terms of these variables the jet masses are $M_{12}^2=Q^2(1-x_3)$, etc.
Then the final result for the subtracted, massless 3-parton cross section
is
\begin{eqnarray}
\label{eq:sigma.3}
  d\hat{\sigma}_3 &=&
  K dL(\hat{l}_1, \hat{l}_2, \hat{l}_3 ;q, 0)
  \frac{g^2}{Q^4}
    \left[
       \left( \frac{1}{1-x_1} + \frac{1}{1-x_2} + \frac{1}{1-x_3}
       \right)^2
    \right.
\nonumber\\
&&\hspace*{1.5cm}
    \left.
       - \frac{ \Theta(Q^2(1-x_1)/\mu_F^2) }{x_1^3(1-x_1)^2}
       - \frac{ \Theta(Q^2(1-x_2)/\mu_F^2) }{x_2^3(1-x_2)^2}
       - \frac{ \Theta(Q^2(1-x_3)/\mu_F^2) }{x_3^3(1-x_3)^2}
    \right] 
\nonumber\\
&& + O(g^4).  
\end{eqnarray}
The factors of $x_i^3$ in the denominators of the subtraction terms result
from the Jacobians in the transformation between three-parton and massless
two-parton kinematics, as explained in Sec.\ \ref{sec:examples}.


%% file: phi3-conclusions.tex

\section{Conclusions}
\label{sec:conclusions}

I have constructed a Monte-Carlo algorithm for $e^+e^-$ annihilation in a
model theory that can be applied at arbitrarily non-leading order (and
hence includes arbitrarily non-leading logarithms).  The main features
are:
\begin{itemize}
\item Corrections to the hard-scattering cross section and to the
  kernel for parton showering are defined by a subtractive method to all
  orders of perturbation theory.
\item The subtractions are applied point-by-point in momentum space, so
  that well-behaved functions are obtained that can be used in the
  calculation of arbitrary observables, without any restriction to
  infra-red-safe quantities.
\item A modified LSZ prescription must be used for the normalization of
  the hard scattering coefficient (and the corresponding parton showering
  kernels).  The normal LSZ prescription requires the on-shell amputated
  amplitudes to be multiplied by an appropriate power of the residue of
  the external parton propagator.  Instead the on-shell amputated and
  subtracted amplitudes are to be multiplied by an appropriate power of an
  integrated jet factor.  
\item Exact momentum conservation is enforced at each stage, with a
  definite prescription for relating exact parton momenta to the
  corresponding approximated massless momenta that are used in the
  calculation of the hard-scattering and splitting coefficients.
\item The produced jet masses are restricted by a cut-off function, and
  there is a renormalization-group-like invariance under changes of the
  cut-off function.
\item The showering algorithm is rather different from a conventional
  showering algorithm.  It is applied in the rest frame of the
  generated off-shell parton.  The interpretation purely in terms of
  DGLAP kernels does not appear to hold beyond the leading-logarithm
  level.
\end{itemize}
Proofs were given that the errors are power suppressed.  Thus the accuracy
of the Monte-Carlo calculation is completely equivalent to that of
conventional analytic calculations of inclusive processes, and so all the
analytic (or matrix-element) results can be used, with a change of scheme,
but now with the advantage of being able to treat arbitrary final-states.
The method of proof appears to be capable of considerable generalization.

The use of a subtractive method to construct the corrections means that
the subtracted higher-order corrections are not necessarily positive.  So
the algorithm must generally be used to generate weighted events.
However, as explained in Ref.\ \cite{MC.BGF}, adjustment of the cut-off
function can be used to reduce the number of negative weighted events.

Of course, a model field theory is not QCD, and future work should aim at
extending the results to QCD, as well as to lepto-production and
hadro-production.  The main problem here is well-known: that soft-gluon
emission occurs at the leading power in any gauge theory, and the
treatment of the leading regions needs to be considerable generalized.
Unlike the case of inclusive cross sections, one cannot appeal directly to
a cancellation of soft-gluon effects between real and virtual emission.
One symptom of this problem is in the definition of the jet factor $Z$.
In our model field theory it is given by a simple cut Green function of
the parton field.  But in a gauge theory, such a definition is not gauge
invariant.  

Collins and Hautmann \cite{Collins:2000dz,Collins:2000gd} have proposed
some ideas at the one-loop level for how these subtractions should be
performed in a way that allows gauge-invariant definitions of the
jet factors and the similar initial-state parton densities.  These
definitions would represent a likely improvement on those by Collins and
Soper \cite{kt2}.  

Of course, coherent angular-ordered showering, such as is implemented in
HERWIG, solves the problem of soft-gluon emission at the leading-logarithm
level and somewhat beyond, but it does not represent a complete solution
to the problem of obtain non-leading corrections in general.
